\documentclass[a4paper]{article}%
\usepackage{amssymb}
\usepackage{amsmath}%
\usepackage{amsfonts}%
\usepackage{graphicx}

\begin{document}

\author{Sawa Manoff\\\textit{Bulgarian Academy of Sciences,}\\\textit{Institute for Nuclear Research and Nuclear Energy,}\\\textit{Department of Theoretical Physics,}\\\textit{Blvd. Tzarigradsko Chaussee 72,}\\\textit{1784 Sofia - Bulgaria}}
\date{E-mail address: smanov@inrne.bas.bg}
\title{\textbf{Null vector fields in spaces with affine connectons and metrics.
Doppler's effect, Hublle's effect, and aberration's effect.}}
\maketitle

\begin{abstract}
\textit{The notion of null (isotropic) vector field is considered in spaces
with affine connections and metrics [}$(\overline{L}_{n},g)$\textit{-spaces]
as models of space or space-time. On its basis the propagation of signals in
space-time is considered. The Doppler effect is generalized for these type of
spaces. The notions of standard Doppler effect and transversal Doppler effect
are introduced. On their grounds, the Hubble effect and the aberration effect
appear as Doppler effects with explicit forms of the centrifugal (centripetal)
and Coriolis velocity vector fields in spaces with affine connections and
metrics. The upper limit of the value of the general observed shift parameter
}$z$\textit{, generated by both the effects, based on the Doppler effects, is
found to be }$z=\sqrt{2}$\textit{. Doppler effects, Hubble's effect, and
aberration's effect could be used in mechanics of continuous media and in
other classical field theories in the same way as the standard Doppler effect
is used in classical and relativistic mechanics.}


\end{abstract}

\section{Introduction}

The notion of null (isotropic) vector field is related to the light
propagation described in relativistic electrodynamics on the basis of special
and general relativity theories \cite{Stephani} $\div$ \cite{Misner}. On the
other side, the notion of null (isotropic) vector field could be considered in
spaces with (definite) or indefinite metric as a geometric object
(contravariant vector field) with specific properties making it useful in the
description of the propagation of signals in space or in space-time as well as
in geometrical optics based on different mathematical models. Usually, it is
assumed that a signal is propagating with limited velocity through a
continuous media or in vacuum. The velocity of propagation of signals could be
a constant quantity or a non-constant quantity depending on the properties of
the space or the space-time, where the signals are transmitted and propagated.
Recently, it has been shown that a classical field theory could be considered
as a theory of continuous media with its kinematic and dynamic characteristic
\cite{Manoff-1} $\div$ \cite{Manoff-2a}. On this basis the propagation of
signals in different models of space or of space-time is worth being investigated.

In the present paper the notion of contravariant null (isotropic) vector field
is introduced and considered in spaces with affine connections and metrics
[$(\overline{L}_{n},g)$-spaces]. In Section 2 the properties of null vector
fields are considered on the basis of $(n-1)+1$ representation of non-null
(non-isotropic) vector fields orthogonal to each other. In Section 3 the
notions of distance and space velocity are discussed and their relations to
null vector fields are investigated. In Section 4 the kinematic effects
[longitudinal and transversal Doppler's effects, Hubble's effect, and
aberration's effect] related to the kinematic characteristics of the relative
velocity and their connections with null vector fields are considered. It is
shown that the Hubble effect and the aberration effect appear as corollaries
of the standard (longitudinal) and transversal Doppler effects. On the other
side, the Hubble effect and the aberration effect are closely related to the
centrifugal (centripetal) and the Coriolis velocities. The results discussed
in the paper could be important from the point of view of the possible
applications of kinematic characteristics in continuous media mechanics as
well as in classical (non-quantum) field theories in spaces with affine
connections and metrics.

The main results in the paper are given in details (even in full details) for
these readers who are not familiar with the considered problems. The
definitions and abbreviations are identical to those used in \cite{Manoff-2}
and \cite{Manoff-2a}. The reader is kindly asked to refer to them for more
details and explanations of the statements and results only cited in this paper.

\section{Null (isotropic) vector fields. Definition and properties}

\subsection{Definition of a null (isotropic) vector field}

Let us now consider a space with affine connections and metrics [$(\overline
{L}_{n},g)$-space] \cite{Manoff-3}, \cite{Manoff-3a} as a model of a space or
of a space-time. In this space the length $l_{v}$ of a contravariant vector
field $v\in T(M)$ is defined by the use of the covariant metric tensor field
(covariant metric) $g\in\otimes_{2s}(M)$ as
\begin{equation}
g(v,v)=\pm l_{v}^{2}\,\,\,\,\,\,\text{,\thinspace\thinspace\thinspace
\thinspace\thinspace\thinspace\thinspace\thinspace\thinspace\thinspace
\thinspace\thinspace\thinspace\thinspace\thinspace\thinspace}l_{v}^{2}%
\geq0\text{ \thinspace.}\label{1.1}%
\end{equation}

\textit{Remark}. The sign before $l_{v}^{2}$ depends on the signature $Sgn$ of
the covariant metric $g$.

The contravariant vector fields can be divided into two classes with respect
to their lengths:

\begin{itemize}
\item null or isotropic vector fields with length $l_{v}=0$,

\item non-null or non-isotropic vector fields with length $l_{v}\neq0$.
\end{itemize}

In the case of a positive definite covariant metric $g$ ($Sgng=\pm n$,
$dimM=n$) the null (isotropic) vector field is identically equal to zero, i.e.
if $l_{v}=0$ then $v=v^{i}\cdot e_{i}\equiv\mathbf{0}\in T(M)$,\thinspace
\thinspace$v^{i}\equiv0$.

In the case of an indefinite covariant metric $g$ ($Sgng<n$ or $Sgng>-n$,
$dimM=n$) the null (isotropic) vector field with equal to zero length
$l_{v}=0$ can have different from zero components in an arbitrary given basis,
i.e. it is not identically equal to zero at the points, where it has been
defined, i.e. if $l_{v}=0$ then $v\neq\mathbf{0}\in T(M)$, $v=v^{i}\cdot
e_{i}\in T(M)$ and $v^{i}\neq0$. In a $(\overline{L}_{n},g)$-space the
components $g_{ij}$ of a covariant metric tensor $g$ could be written in a
local co-ordinate system at a given point of the space as $g_{ij}%
=(\underset{k\,\,\,times}{\underbrace{-1,\,-1,\,-1,\,...,}}\,\underset
{l\,\,\,\,\,times}{\underbrace{+1,\,+1,\,+1,\,...}})$ with $k+l=n$.

The signature $Sgn$ of $g$ is defined as
\begin{equation}
Sgn\,\,g=-k+l=2\cdot l-n=n-2\cdot k\,\,\,\,\text{, \thinspace\thinspace
\thinspace\thinspace}n,\,k,\,l\in\mathbf{N}\text{,}\label{1.2}%
\end{equation}
\thinspace\thinspace\thinspace where $k=n-l$,\thinspace\thinspace
\thinspace$l=n-k$.

In the relativistic physics for $dim\,M=4$, the number $l$ and $k$ are chosen
as $l=1$, $k=3$ or $l=3$, $k=1$ so that $Sgn\,g=-2\sim$ $(-1,\,-1,\,-1,\,+1)$
or $Sgn\,g=+2$ $\sim(+1,\,+1,\,+1,\,-1)$. In general, a $(\overline{L}_{n}%
,g)$-space could be consider as a model of space-time with $Sgn\,g<0$ and
$(k>l$, $l=1)$ or with $Sgn\,g>0$ and $(l>k$, $k=1)$.

The non-null (non-isotropic) contravariant vector fields are divided into two classes.

1. For $Sgn\,g<0$

(a) $g(v,v)=+l_{v}^{2}>0$ $:=$ time like vector field $v\in T(M)$,

(b) $g(v,v)=-l_{v}^{2}<0$ $:=$ space like vector field $v\in T(M)$.

2. For $sgn\,g>0$

(a) $g(v,v)=-l_{v}^{2}<0$ $:=$ time like vector field $v\in T(M)$,

(b) $g(v,v)=+l_{v}^{2}>0\,:=$ space like vector field $v\in T(M)$.

Therefore, if we do not fix a priory the signature of the space-time models we
can distinguish a \textit{time like vector field} $u$ with

\begin{center}%
\begin{tabular}
[c]{ll}%
$g(u,u)=+l_{u}^{2}\text{ \thinspace\thinspace\thinspace\thinspace\thinspace
for}$ & $Sgn\,g<0$\\
$\,\,\,\,\,\,\,\,\,\,\,\,\,\,\,\,\,\,\,=-l_{u}^{2}\,\,\,\,\,\,$for &
$Sgn\,g>0$%
\end{tabular}

\end{center}

or $g(u,u)=\pm l_{u}^{2}$, and a \textit{space like vector field} $\xi_{\perp
}$ with

\begin{center}%
\begin{tabular}
[c]{ll}%
$g(\xi_{\perp},\xi_{\perp})=-l_{\xi_{\perp}}^{2}\text{ \thinspace
\thinspace\thinspace\thinspace\thinspace for}$ & $Sgn\,g<0$\\
$\,\,\,\,\,\,\,\,\,\,\,\,\,\,\,\,\,\,\,=+l_{\xi_{\perp}}^{2}\,\,\,\,\,\,$for &
$Sgn\,g>0$%
\end{tabular}

\end{center}

or $g(\xi_{\perp},\xi_{\perp})=\mp l_{\xi_{\perp}}^{2}$. This means that in
symbols $\pm l_{\diamond}^{2}$ or $\mp l_{\diamond}^{2}$ \thinspace
\thinspace$(\diamond\in T(M))$ the sign above is related to $Sgn\,g<0$ and the
sign below is related to $Sgn\,g>0$.

\textit{Remark}. Since $l_{\diamond}=\pm\sqrt{l_{\diamond}^{2}}$, the sings in
this case will be denoted as not related to the signature of the metric $g $.

A non-null (non-isotropic) contravariant vector field $v$ could be represented
by its length $l_{v}$ and its corresponding unit vector $n_{v}=\frac v{l_{v}}$
as $v=\pm l_{u}\cdot n_{v}$ in contrast to a null (isotropic) vector field
$\widetilde{k}$ with $l_{\widetilde{k}}=0\,$\thinspace\thinspace(the sings
here are not related to the signature of the metric $g$)$\,$%
\[
v=\pm l_{v}\cdot n_{v}\text{ \thinspace\thinspace\thinspace\thinspace
,\thinspace\thinspace\thinspace\thinspace\thinspace\thinspace\thinspace
\thinspace}g(v,v)=l_{v}^{2}\cdot g(n_{v},n_{v})=\pm l_{v}^{2}\text{\thinspace
\thinspace\thinspace\thinspace\thinspace,\thinspace\thinspace\thinspace
\thinspace\thinspace\thinspace\thinspace\thinspace}g(n_{v},n_{v}%
)=\pm1\,\,\,\text{,}
\]
or
\[
v=\pm l_{v}\cdot n_{v}\text{ \thinspace\thinspace\thinspace\thinspace
,\thinspace\thinspace\thinspace\thinspace\thinspace\thinspace\thinspace
\thinspace}g(v,v)=l_{v}^{2}\cdot g(n_{v},n_{v})=\mp l_{v}^{2}\text{\thinspace
\thinspace\thinspace\thinspace\thinspace,\thinspace\thinspace\thinspace
\thinspace\thinspace\thinspace\thinspace\thinspace}g(n_{v},n_{v}%
)=\mp1\,\,\,\text{.}
\]

\textit{Remark}. In the experimental physics, the measurements are related to
the lengths and to the directions of a non-null (non-isotropic) vector field
with respect to a frame of reference. Since different types of co-ordinates
could be used in a frame of reference, the components of a vector field
related to these co-ordinates cannot be considered as invariant
characteristics of the vector field and on this grounds the components cannot
be important characteristics for the vector fields.

After these preliminary remarks, we can introduce the notion of a null
(isotropic) vector field

\textit{Definition 1.} A contravariant vector field $\widetilde{k}$ with
length zero is called null (isotropic) vector field, i.e. $\widetilde{k}:=$
null (isotropic) vector field if
\begin{equation}
g(\widetilde{k},\widetilde{k})=\pm l_{\widetilde{k}}^{2}%
=0\,\,\,\,\,\text{,\thinspace\thinspace\thinspace\thinspace\thinspace
\thinspace\thinspace\thinspace\thinspace\thinspace\thinspace\thinspace
}l_{\widetilde{k}}=\mid g(\widetilde{k},\widetilde{k})\mid^{1/2}%
=0\,\,\,\,\text{.}\label{1.3}%
\end{equation}

\subsection{Properties of a null (isotropic) vector field}

The properties of a null (isotropic) contravariant vector field could be
considered in a $(n-1)+1$ invariant decomposition of a space-time by the use
of two non-isotropic contravariant vector fields $u$ and $\xi_{\perp}$,
orthogonal to each other \cite{Manoff-3a}, i.e. $g(u,\xi_{\perp})=0$. The
contravariant vectors $u$ and $\xi_{\perp}$ are essential elements of the
structure of a frame of reference \cite{Manoff-4} in a space-time.

\subsubsection{Invariant representation of a null vector field by the use of a
non-null contravariant vector field}

(a) \textit{Invariant projections of a null vector field along and orthogonal
to a non-null (non-isotropic) contravariant vector field} $u$

Every contravariant vector field $\widetilde{k}\in T(M)$ could be represented
in the form
\begin{equation}
\widetilde{k}=\frac1e\cdot g(u,\widetilde{k})\cdot u+\overline{g}%
[h_{u}(\widetilde{k})]=k_{\parallel}+k_{\perp}\,\,\,\,\text{,}\label{1.4}%
\end{equation}
where
\begin{align}
e  & =g(u,u)=\pm l_{u}^{2}\text{ \thinspace\thinspace\thinspace\thinspace
,\thinspace\thinspace\thinspace\thinspace\thinspace\thinspace\thinspace
\thinspace\thinspace}\nonumber\\
\overline{g}  & =g^{ij}\cdot\partial_{i}.\partial_{j}%
\,\,\,\,\,\text{,\thinspace\thinspace\thinspace\thinspace\thinspace
\thinspace\thinspace\thinspace\thinspace\thinspace}\partial_{i}.\partial
_{j}=\frac12\cdot(\partial_{i}\otimes\partial_{j}+\partial_{j}\otimes
\partial_{i})\,\,\,\,\text{,}\nonumber\\
g  & =g_{ij}\cdot dx^{i}.dx^{j}\,\,\,\,\text{,\thinspace\thinspace\thinspace
}dx^{i}.dx^{j}=\frac12\cdot(dx^{i}\otimes dx^{j}+dx^{j}\otimes dx^{i})\text{
\thinspace,}\nonumber\\
h_{u}  & =g-\frac1e\cdot g(u)\otimes g(u)\,\,\,\,\,\text{,\thinspace
\thinspace\thinspace\thinspace\thinspace}h^{u}=\overline{g}-\frac1e\cdot
u\otimes u\,\,\,\,\,\text{,}\nonumber\\
g(u)  & =g_{i\overline{j}}\cdot u^{j}\cdot dx^{i}\,\,\,\text{,\thinspace
\thinspace\thinspace}\nonumber\\
\text{\thinspace\thinspace\thinspace\thinspace}\overline{g}[h_{u}%
(\widetilde{k})]  & =g^{ij}\cdot h_{\overline{j}\overline{l}}\cdot
\widetilde{k}\,^{l}\cdot\partial_{i}:=k_{\perp}\,\,\,\text{,\thinspace
\thinspace\thinspace\thinspace\thinspace\thinspace}k_{\parallel}:=\frac1e\cdot
g(u,\widetilde{k})\cdot u\,\,\,\,\text{.}\label{1.5}%
\end{align}
\begin{equation}
g(k_{\parallel},k_{\perp})=0\,\,\,\,\,\,\text{,\thinspace\thinspace
\thinspace\thinspace\thinspace\thinspace\thinspace\thinspace\thinspace
\thinspace\thinspace\thinspace}g(u,k_{\perp})=0\,\,\,\,\,\,\,\text{.}%
\label{1.6}%
\end{equation}

Let us now take a closer look at the first term $k_{\parallel}$ of the
representation of $\widetilde{k}$.
\begin{equation}
k_{\parallel}=\frac1e\cdot g(u,\widetilde{k})\cdot u=\pm\frac1{l_{u}^{2}}\cdot
g(u,\widetilde{k})\cdot u=\pm\frac1{l_{u}}\cdot g(u,k_{\parallel})\cdot\frac
u{l_{u}}\text{ .}\label{1.7}%
\end{equation}

If we introduce the abbreviations
\begin{equation}
n_{\parallel}=\frac u{l_{u}}\text{ \thinspace\thinspace,\thinspace
\thinspace\thinspace\thinspace\thinspace\thinspace\thinspace\thinspace}%
\omega=g(u,\widetilde{k})=g(u,k_{\parallel})\,\,\,\,\,\,\,\text{,}\label{1.8}%
\end{equation}
where
\begin{equation}
g(n_{\parallel},n_{\parallel})=\frac1{l_{u}^{2}}\cdot g(u,u)=\frac1{l_{u}^{2}%
}\cdot(\pm l_{u}^{2})=\pm1\text{ \thinspace\thinspace\thinspace,}\label{1.9}%
\end{equation}
\begin{align}
\omega & =g(u,\widetilde{k})=g(u,k_{\parallel}+k_{\perp})=g(u,k_{\parallel
})=l_{u}\cdot g(n_{\parallel},k_{\parallel})=l_{u}\cdot g(k_{\parallel
},n_{\parallel})\,\,\,\text{,}\label{1.10}\\
k_{\parallel}  & :=\,\pm l_{k_{\parallel}}\cdot n_{\parallel}%
\,\,\,\,\,\text{,\thinspace\thinspace\thinspace\thinspace\thinspace
\thinspace\thinspace\thinspace\thinspace}g(k_{\parallel},k_{\parallel
})=l_{k_{\parallel}}^{2}\cdot g(n_{\parallel},n_{\parallel})=l_{k_{\parallel}%
}^{2}\cdot(\pm1)=\pm l_{k_{\parallel}}^{2}\,\,\,\text{,}\label{1.10a}\\
g(k_{\parallel},n_{\parallel})  & =\pm\,l_{k_{\parallel}}\cdot g(n_{\parallel
},n_{\parallel})=\pm\,l_{k_{\parallel}}\cdot(\pm1)=l_{k_{\parallel}}%
=\frac\omega{l_{u}}\,\,\,\text{,}\label{1.10b}%
\end{align}
then $k_{\parallel}$ could be expressed as (the signs are not related to the
signature of the metric $g$)
\begin{equation}
k_{\parallel}=\pm\frac\omega{l_{u}}\cdot n_{\parallel}\,=\pm\,l_{k_{\parallel
}}\cdot n_{\parallel}\,\,\,\,\text{\thinspace\thinspace.}\label{1.11}%
\end{equation}

The vector $n_{\parallel}$ is a unit vector $[g(n_{\parallel},n_{\parallel
})=\pm1]$ collinear to $u$ and, therefore, tangential to a curve with
parameter $\tau$ if $u=\frac d{d\tau}$.

The scalar invariant $\omega=g(u,\widetilde{k})$ is usually interpreted as the
frequency of the radiation related to the null vector field $\widetilde{k}$
and propagating with velocity $u$ with absolute value $l_{u}$ with respect to
the trajectory $x^{i}(\tau)$. In general relativity $l_{u}:=c$ and it is
assumed that the radiation is of electromagnetic nature propagating with the
velocity of light $c$ in vacuum. We will come back to this interpretation in
the next considerations.

The contravariant vector field $k_{\perp}$%
\[
k_{\perp}=\overline{g}[h_{u}(\widetilde{k})]
\]
is orthogonal to $u$ (and $k_{\parallel}$ respectively) part of $\widetilde
{k}$. Since
\begin{align}
g(k_{\parallel},k_{\parallel})  & =g(\pm\frac\omega{l_{u}}\cdot n_{\parallel
},\,\pm\frac\omega{l_{u}}\cdot n_{\parallel})=\frac{\omega^{2}}{l_{u}^{2}%
}\cdot g(n_{\parallel},n_{\parallel})=\pm\frac{\omega^{2}}{l_{u}^{2}}=\pm
l_{k_{\parallel}}^{2}\,\,\,\,\,\text{,}\label{1.12}\\
l_{k_{\parallel}}  & =\frac\omega{l_{u}}\text{ }>0\text{\thinspace
\thinspace\thinspace\thinspace,\thinspace\thinspace\thinspace\thinspace
\thinspace\thinspace}l_{k_{\parallel}}^{2}=\frac{\omega^{2}}{l_{u}^{2}%
}\,\,\,\,\text{,}\label{1.13}%
\end{align}
and \thinspace%
\[
g(\widetilde{k},\widetilde{k})=0
\]
we have for $g(k_{\perp},k_{\perp})$%
\begin{align}
g(\widetilde{k},\widetilde{k})  & =0=g(k_{\parallel}+k_{\perp},\,k_{\parallel
}+k_{\perp})=g(k_{\parallel},k_{\parallel})+g(k_{\perp},k_{\perp})=\nonumber\\
& =\pm\frac{\omega^{2}}{l_{u}^{2}}+g(k_{\perp},k_{\perp})=\pm\frac{\omega^{2}%
}{l_{u}^{2}}\mp l_{k_{\perp}}^{2}=0\,\,\,\,\,\text{,}\label{1.15}%
\end{align}
and, therefore,
\begin{equation}
l_{k_{\perp}}^{2}=\frac{\omega^{2}}{l_{u}^{2}}%
\,\,\,\,\,\,\,\,\text{,\thinspace\thinspace\thinspace\thinspace\thinspace
\thinspace\thinspace\thinspace\thinspace\thinspace\thinspace\thinspace
}l_{k_{\perp}}=\frac\omega{l_{u}}=l_{k_{\parallel}}\,\,\,\,\text{.}%
\label{1.16}%
\end{equation}

\textit{Remark}. Since $\omega\geq0$ and $l_{u}>0$, and at the same time
$l_{k_{\perp}}>0$, and $l_{k_{\parallel}}>0$, we have
\[
l_{k_{\parallel}}=\frac\omega{l_{u}}=l_{k_{\perp}}\,\,\,\,\text{.}
\]

If we introduce the unit contravariant vector $\widetilde{n}_{\perp}$ with
$g(\widetilde{n}_{\perp},\widetilde{n})=\mp1$ then the vector $k_{\perp}$
could be written as
\begin{equation}
k_{\perp}:=\mp\,l_{k_{\perp}}\cdot\widetilde{n}_{\perp}\,\,\,\,\,\,\text{,}%
\label{1.17}%
\end{equation}
where
\begin{equation}
g(k_{\perp},k_{\perp})=l_{k_{\perp}}^{2}\cdot g(\widetilde{n}_{\perp
},\widetilde{n})=\mp l_{k_{\perp}}^{2}=\mp\,\frac{\omega^{2}}{l_{u}^{2}%
}\,\,\,\,\text{, \thinspace\thinspace\thinspace\thinspace\thinspace\thinspace
}l_{k_{\perp}}^{2}=\frac{\omega^{2}}{l_{u}^{2}}\,\,\,\,\,\text{,\thinspace
\thinspace\thinspace\thinspace\thinspace\thinspace\thinspace}l_{k_{\perp}%
}=\frac\omega{l_{u}}\,\,\text{.}\label{1.18}%
\end{equation}

Therefore,
\begin{align}
k_{\perp}  & =\mp\frac\omega{l_{u}}\cdot\widetilde{n}_{\perp}%
\,\,\,\,\,\,\text{,\thinspace\thinspace\thinspace\thinspace\thinspace
\thinspace\thinspace\thinspace\thinspace\thinspace\thinspace\thinspace
\thinspace\thinspace}k_{\parallel}=\pm\frac\omega{l_{u}}\cdot n_{\parallel
}\,\,\,\,\,\,\,\text{,}\label{1.19}\\
k  & =k_{\parallel}+k_{\perp}=\pm\frac\omega{l_{u}}\cdot(n_{\parallel
}-\widetilde{n}_{\perp})\,\,\,\,\,\,\text{,}\label{1.20}%
\end{align}
where
\begin{equation}
g(n_{\parallel},\widetilde{n}_{\perp})=0\,\,\,\,\,\text{,\thinspace
\thinspace\thinspace\thinspace\thinspace\thinspace\thinspace\thinspace
}g(k_{\parallel},\xi_{\perp})=0\,\,\,\,\,\text{,\thinspace\thinspace
\thinspace\thinspace\thinspace\thinspace\thinspace\thinspace\thinspace
}g(u,k_{\perp})=0\,\,\,\,\,\,\text{,}\label{1.21}%
\end{equation}
\begin{align}
g(n_{\parallel},k_{\parallel})  & =\pm l_{k_{\parallel}}\cdot g(n_{\parallel
},n_{\parallel})=l_{k_{\perp}}=\frac\omega{l_{u}}\,\,\,\,\,\,\text{,}%
\label{1.22}\\
g(\widetilde{n}_{\perp},k_{\perp})  & =\mp l_{k_{\perp}}\cdot g(n_{\perp
},n_{\perp})=\frac\omega{l_{u}}\,\,=l_{k_{\parallel}}=l_{k_{\perp}%
}\,\,\,\,\,\text{.}\label{1.23}%
\end{align}
\[
g(n_{\parallel},k_{\parallel})=g(\widetilde{n}_{\perp},k_{\perp})=\frac
\omega{l_{u}}=l_{k_{\parallel}}=l_{k_{\perp}}\,\,\,\,\,\,\text{.}
\]

\textit{Remark}. The signs not related to the metric $g$ are chosen so to be
the same with the signs related to the metric $g$.

We have now the relations
\begin{equation}
\omega=g(u,\widetilde{k})=l_{u}\cdot g(n_{\parallel},k_{\parallel})=l_{u}\cdot
g(\widetilde{n}_{\perp},k_{\perp})\,\,\,\,\,\text{.}\label{1.24}%
\end{equation}

If $\widetilde{n}_{\perp}$ is interpreted as the unit vector in the direction
of the propagation of a signal in the subspace orthogonal to the contravariant
vector field $u$ and $l_{u}$ is interpreted as the absolute value of the
velocity of the radiated signal then $l_{u}\cdot\widetilde{n}_{\perp}$ is the
path along $\widetilde{n}_{\perp}$ propagated by the signal in a unit time
interval. Then
\begin{equation}
\omega=g(u_{\perp},k_{\perp})\text{ \thinspace\thinspace\thinspace
\thinspace,\thinspace\thinspace\thinspace\thinspace\thinspace\thinspace
\thinspace\thinspace}u_{\perp}:=l_{u}\cdot\widetilde{n}_{\perp}%
\,\,\,\,\,\,\text{,\thinspace\thinspace\thinspace\thinspace\thinspace
\thinspace\thinspace\thinspace\thinspace\thinspace}g(u,u_{\perp}%
)=0\,\,\,\,\,\text{.}\label{1.25}%
\end{equation}

Let us now consider more closely the explicit form of $k_{\perp}$%
\[
k_{\perp}=\mp l_{k_{\perp}}\cdot\widetilde{n}_{\perp}=\mp\frac\omega{l_{u}%
}\cdot\widetilde{n}_{\perp}\text{ \thinspace\thinspace\thinspace.}
\]

(a) In $3$-dimensional Euclidean space (as model of space-time of the
Newtonian mechanics) the wave vector $\overrightarrow{k}$ is defined as
\[
\overrightarrow{k}=\frac{2\pi}\lambda\cdot\overrightarrow{n}\,\,\,\text{,}
\]
where $\overrightarrow{n}$ is the unit $3$-vector in the direction of
propagation of a signal with absolute value of its velocity $l_{u}%
=\lambda\cdot\nu$. If we express $\lambda$ by $\lambda=l_{u}/\nu$ and put the
equivalent expression in this for $\overrightarrow{k}$ we obtain the
expression
\[
\overrightarrow{k}=\frac{2\pi\cdot\nu}{l_{u}}\cdot\overrightarrow{n}%
=\frac\omega{l_{u}}\cdot\overrightarrow{n}\,\,\,\,\text{,}
\]
which (up to a sign depending on the signature of the metric $g$) is identical
with the expression for $k_{\perp}$ for $n=3$ if $k_{\perp}=\overrightarrow
{k}$, $\overrightarrow{n}=\widetilde{n}_{\perp}$,\thinspace and $\omega
=2\cdot\pi\cdot\nu$.

(b) In $4$-dimensional (pseudo) Riemannian space (as a model of space-time of
the Einstein theory of gravitation) $l_{u}$ is interpreted as the absolute
value of the velocity of light in vacuum (normalized by some authors to $1$),
i.e. $l_{u}=c$, $1$. Then
\[
k_{\perp}=\mp\frac\omega c\cdot\widetilde{n}_{\perp}=\mp\frac{2\cdot\pi
\cdot\nu}{\lambda\cdot\nu}\cdot\widetilde{n}_{\perp}=\mp\frac{2\cdot\pi
}\lambda\cdot\widetilde{n}_{\perp}
\]
and we obtain the expression for the wave vector of light propagation in
general relativity, where $\widetilde{n}_{\perp}$ is the unit vector along the
propagation of light in the corresponding $3$-dimensional subspace of an
observer with world line $x^{i}(\tau)$ if
\[
u=\frac d{d\tau}=l_{u}\cdot n_{\parallel}\,\,\,\,\,\,\,\text{,\thinspace
\thinspace\thinspace\thinspace\thinspace\thinspace\thinspace\thinspace
\thinspace\thinspace\thinspace}\,n_{\parallel}=\frac1{l_{u}}\cdot\frac
d{d\tau}\text{ \thinspace\thinspace\thinspace.}
\]

$l_{u}$ is the velocity of light measured by the observer.

(c) In the general case for $k_{\perp}$ as
\[
k_{\perp}=\mp\frac\omega{l_{u}}\cdot\widetilde{n}_{\perp}%
\,\,\,\text{\thinspace\thinspace\thinspace\thinspace\thinspace,}
\]
$\omega$ could also be interpreted as the frequency of a signal propagating
with velocity with absolute value $l_{u}$ in a frame of reference of an
observer with world line $x^{i}(\tau)$. The unit vector $\widetilde{n}_{\perp
}$ is the unit vector in the direction of the propagation of the signal in the
subspace orthogonal to the vector $u$. The velocity of the observer is usually
defined by the use of the parameter $\tau$ of the world line under the
assumption that $ds=l_{u}\cdot d\tau$, where $ds$ is the distance of the
propagation of a signal for the proper time interval $d\tau$ of the observer
\[
u=\frac d{d\tau}=\frac d{\frac1{l_{u}}\cdot ds}=l_{u}\cdot\frac d{ds}%
\,\,\,\text{.}
\]

\textit{Remark}. Usually the velocity of a particle (observer) moving in
space-time is determined by its velocity vector field $u=\frac d{d\tau}$,
where $\tau$ is the proper time of the observer. The parameter $\tau$ is
considered as a parameter of its world line $x^{i}(\tau)$. By the use of $u$
and its corresponding projection metrics $h_{u}$ and $h^{u}$ a contravariant
(non-null, non-isotropic) vector field $\xi$ could be represented in two
parts: one part is collinear to $u$ and the other part is orthogonal to $u$%
\[
\xi=\frac1e\cdot g(\xi,u)\cdot u+\overline{g}[h_{u}(\xi)]=\xi_{\parallel}%
+\xi_{\perp}\,\,\,\text{,}
\]
where
\[
\xi_{\parallel}=\frac1e\cdot g(\xi,u)\cdot u\,\,\,\,\,\,\,\text{,\thinspace
\thinspace\thinspace\thinspace\thinspace\thinspace\thinspace\thinspace
\thinspace\thinspace}\xi_{\perp}=\text{\thinspace}\overline{g}[h_{u}%
(\xi)]\,\,\,\,\,\text{,\thinspace\thinspace\thinspace\thinspace\thinspace
\thinspace\thinspace\thinspace\thinspace\thinspace\thinspace\thinspace}%
g(\xi_{\parallel},\,\xi_{\perp})=0\,\,\,\,\text{.}
\]

1. If an observer is moving with velocity $v=\frac d{d\overline{\tau}}$ on its
world line $x^{i}(\overline{\tau})$ then its velocity, considered with respect
to the observer with velocity $u$ and world line $x^{i}(\tau)$, will have two
parts $v_{\parallel}$ and $v_{\perp}$, collinear and orthogonal to $u$
$\,$respectively at the cross point $\tau=\overline{\tau}$ of both the world
lines $x^{i}(\tau)$ and $x^{i}(\overline{\tau})$%
\[
v=\frac1e\cdot g(v,u)\cdot u+\overline{g}[h_{u}(v)]=v_{\parallel}+v_{\perp
}\,\,\,\,\,\,\text{.}
\]

The vector $v_{\parallel}$ describes the motion of the observer with velocity
$v$ along the world line of the first observer with velocity $u$. The vector
$v_{\perp}$ describes the motion of the second observer with velocity $v$ in
direction orthogonal to the world line of the first observer. The vector
$v_{\perp}$ is the velocity of the second observer in the space of the first
observer in contrast to the vector $v_{\parallel}$ describing the change of
$v$ in the time of the first observer.

2. If we consider the propagation of a signal characterized by its null vector
field $\widetilde{k}$ the interpretation of the vector field $u$ tangential to
the world line of an observer changes. The vector field $u=l_{u}\cdot
n_{\parallel}$ is interpreted as the velocity vector field of the signal,
propagating in the space-time and measured by the observer at its world line
$x^{i}(\tau)$ with proper time $\tau$ as a parameter of this world line. The
absolute value $l_{u}$ of $u$ is the size of the velocity of the signal
measured along the unit vector field $n_{\parallel}$ collinear to the tangent
vector of the world line of the observer.

3. In Einstein's theory of gravitation (ETG) both interpretations of the
vector field $u$ are put together. On the one side, the vector field $u$ is
interpreted as the velocity of an observer on its world line with parameter
$\tau$ interpreted as the proper time of the observer. On the other side, the
length $l_{u}$ of the vector field $u$ is normalized either to $\pm1$ or to
$\pm c=$ const. The quantity $c$ is interpreted as the light velocity in
vacuum. The basic reason for this normalization is the possibility for
normalization of every non-null (non-isotropic) vector field $u$ in the form
\[
n_{u}=\frac u{l_{u}}=n_{\parallel}\text{ \thinspace\thinspace\thinspace
\thinspace\thinspace,\thinspace\thinspace\thinspace\thinspace where\thinspace
\thinspace\thinspace\thinspace\thinspace\thinspace\thinspace\thinspace
\thinspace\thinspace\thinspace\thinspace\thinspace\thinspace}l_{u}=\mid
g(u,u)\mid^{1/2}\neq0\,\,\,\,\,\,\text{,}
\]
by the use of its different from zero length $(l_{u}\neq0)$, defined by means
of the covariant metric tensor $g$.

Both the interpretations of the vector field $u$ (as velocity of an observer
or as velocity of a signal) should be considered separately from each other
for avoiding ambiguities. The identification of the interpretations should
mean that we assume the existence of an observer moving in space-time with
velocity $u$ and emitting (or receiving) signals with the same velocity. Such
assumption does not exist in the Einstein theory of gravitation. This problem
is worth to be investigated and a clear difference between both
interpretations should be found. It is related to the notions of distance and
velocity in spaces with affine connections and metrics.

\section{Distance and velocity in a $(\overline{L}_{n},g)$-space}

\subsection{Distance in a $(\overline{L}_{n},g)$-space and its relations to
the notion of velocity}

1. The distance in a $(\overline{L}_{n},g)$-space between a point $P\in M$
with co-ordinates $x^{i}$ and a point $\overline{P}\in M$ with co-ordinates
$\overline{x}^{i}=x^{i}+dx^{i}$ is determined by the length of the ordinary
differential $d$, considered as a contravariant vector field $d=dx^{i}%
\cdot\partial_{i}$ \cite{Manoff-2}. If we denote the distance as $ds$ between
point $P$ and point $\overline{P}$ then the square $ds^{2}$ of $ds$ could be
defined as the square of the length of the ordinary differential $d$%
\begin{equation}
ds^{2}=g(d,d)=\pm l_{d}^{2}=g_{\overline{i}\overline{j}}\cdot dx^{i}\cdot
dx^{j}\text{ \thinspace\thinspace\thinspace\thinspace\thinspace\thinspace
,\thinspace\thinspace\thinspace\thinspace\thinspace\thinspace\thinspace
\thinspace\thinspace\thinspace\thinspace\thinspace}l_{d}^{2}\geq
0\,\,\,\,\,\,\,\text{.}\label{2.1}%
\end{equation}

2. Let us now consider a two parametric congruence of curves (a set of not
intersecting curves) in a $(\overline{L}_{n},g)$-space
\begin{equation}
x^{i}=x^{i}(\tau,r(\tau,\lambda))=x^{i}(\tau,\lambda)\text{ \thinspace
\thinspace\thinspace,}\label{2.2}%
\end{equation}
where the function $r=r(\tau,\lambda)\in C^{r}(M)$, $r\geq2$, depends on the
two parameters $\tau$ and $\lambda$, $\tau,\lambda\in\mathbf{R}$. Then
\[
dr=\frac{\partial r(\tau,\lambda)}{\partial\tau}\cdot d\tau+\frac{\partial
r(\tau,\lambda)}{\partial\lambda}\cdot d\lambda\,\,\,\,\,
\]
and
\begin{align*}
dx^{i}  & =\frac{\partial x^{i}(\tau,r(\tau,\lambda))}{\partial\tau}\cdot
d\tau+\frac{\partial x^{i}(\tau,r(\tau,\lambda))}{\partial r}\cdot
(\frac{\partial r(\tau,\lambda)}{\partial\tau}\cdot d\tau+\frac{\partial
r(\tau,\lambda)}{\partial\lambda}\cdot d\lambda)=\\
& =[\frac{\partial x^{i}(\tau,r(\tau,\lambda))}{\partial\tau}+\frac{\partial
x^{i}(\tau,r(\tau,\lambda))}{\partial r}\cdot\frac{\partial r(\tau,\lambda
)}{\partial\tau}]\cdot d\tau+\\
& +\frac{\partial x^{i}(\tau,r(\tau,\lambda))}{\partial r}\cdot\frac{\partial
r(\tau,\lambda)}{\partial\lambda}\cdot d\lambda
\end{align*}
or
\[
dx^{i}=(u^{i}+\overline{\xi}\,^{i}\cdot l_{v})\cdot d\tau+\overline{\xi}%
\,^{i}\cdot\frac{\partial r}{\partial\lambda}\cdot d\lambda
\,\,\,\,\,\,\,\text{,}
\]
where
\[
u^{i}=\frac{\partial x^{i}(\tau,r(\tau,\lambda))}{\partial\tau}%
\,\,\,\,\,\text{,\thinspace\thinspace\thinspace\thinspace\thinspace
\thinspace\thinspace\thinspace\thinspace\thinspace}l_{v}=\frac{\partial
r(\tau,\lambda)}{\partial\tau}\,\,\,\,\,\,\,\text{,\thinspace\thinspace
\thinspace\thinspace\thinspace\thinspace\thinspace\thinspace\thinspace
}\overline{\xi}\,^{i}=\frac{\partial x^{i}(\tau,r(\tau,\lambda))}{\partial
r}\,\,\,\,\,\,\text{,\thinspace\thinspace}
\]
and
\begin{align*}
d  & =dx^{i}\cdot\partial_{i}=d\tau\cdot(u+l_{v}\cdot\overline{\xi})+(\partial
r/\partial\lambda)\cdot d\lambda\cdot\overline{\xi}\,\,\,\,\,\,\text{,}\\
\frac{d\lambda}{d\tau}  & =0\,\,\,\,\,\text{,\thinspace\thinspace
\thinspace\thinspace\thinspace\thinspace\thinspace\thinspace}\frac{d\tau
}{d\lambda}=0\,\,\,\,\,\text{.}%
\end{align*}

The change of the contravariant vector field $d$ under the change $d\tau$ of
the parameter $\tau$ could be expressed in the form
\[
\frac d{d\tau}=\frac{dx^{i}}{d\tau}\cdot\partial_{i}=u+l_{v}\cdot\overline
{\xi}=\overline{u}\,^{i}\cdot\partial_{i}=\overline{u}%
\,\,\,\,\,\text{,\thinspace\thinspace\thinspace\thinspace\thinspace
\thinspace\thinspace\thinspace\thinspace}\overline{u}^{i}=\frac{dx^{i}}{d\tau
}\,\,\,\,\,\text{,}
\]
where the relations are valid
\begin{align*}
g(\overline{u},u)  & =g(u,u)+l_{v}\cdot g(\overline{\xi},u)\,\,\,\,\text{,}\\
g(\overline{u},\overline{\xi})  & =g(u,\overline{\xi})+l_{v}\cdot
g(\overline{\xi},\overline{\xi})\,\,\,\,\,\text{,}\\
g(\overline{u},\overline{u})  & =g(u+l_{v}\cdot\overline{\xi},u+l_{v}%
\cdot\overline{\xi})=\\
& =g(u,u)+2\cdot l_{v}\cdot g(u,\overline{\xi})+l_{v}^{2}\cdot g(\overline
{\xi},\overline{\xi})\text{ .}%
\end{align*}

The contravariant vector field $\overline{u}=\overline{u}^{i}\cdot\partial
_{i}$ is usually interpreted as the velocity of an observer moving in a
space-time described by a $(\overline{L}_{n},g)$-space as its model. The
contravariant vector $u$ is a tangent vector field to the curve $x^{i}%
(\tau,r(\tau,\lambda)=r_{0}=$ const.$)=x^{i}(\tau,\lambda=\lambda_{0}=$
const.$)$%
\begin{align*}
u  & =u^{i}\cdot\partial_{i}=\frac{\partial x^{i}}{\partial\tau}\cdot
\partial_{i}\,\,\,\,\text{,}\\
\overline{u}  & =\frac1{g(u,u)}\cdot g(\overline{u},u)\cdot u+\overline
{g}[h_{u}(\overline{u})]\,\,\,\,\text{.}%
\end{align*}

The contravariant vector $\overline{\xi}$ is a collinear vector to the tangent
vector $\xi$ to the curve $x^{i}(\tau=\tau_{0}=$ const.$,\,r(\tau_{0}%
,\lambda))=x^{i}(\tau=\tau_{0}=$ const.$,\lambda)$. Since
\[
\overline{\xi}=\overline{\xi}\,^{i}\cdot\partial_{i}=\frac{\partial x^{i}%
}{\partial r}\cdot\partial_{i}\,\,\,\,\,
\]
then
\[
\frac{\partial x^{i}}{\partial r}=\frac{\partial x^{i}(\tau,r(\tau,\lambda
))}{\partial r}=\frac{\partial x^{i}(\tau,\lambda(r,\tau))}{\partial r}%
=\frac{\partial x^{i}}{\partial\lambda}\cdot\frac{\partial\lambda}{\partial
r}=\xi^{i}\cdot\frac{\partial\lambda}{\partial r}=\overline{\xi}^{i}\text{
\thinspace,}
\]
where
\begin{align*}
r  & =r(\lambda,\tau)\,\,\,\,\text{,\thinspace\thinspace\thinspace
\thinspace\thinspace}\lambda=\lambda(\tau,r)\text{ \thinspace\thinspace
\thinspace\thinspace\thinspace,}\\
\overline{\xi}  & =\overline{\xi}\,^{i}\cdot\partial_{i}=\frac{\partial
\lambda}{\partial r}\cdot\xi^{i}\cdot\partial_{i}=\frac{\partial\lambda
}{\partial r}\cdot\xi\,\,\,\,\text{,\thinspace\thinspace\thinspace
\thinspace\thinspace\thinspace\thinspace\thinspace\thinspace}\xi
=\frac{\partial x^{i}}{\partial\lambda}\cdot\partial_{i}\,\,\,\,\text{.}%
\end{align*}

3. Further, since we wish to consider the vector field $u$ as the velocity
vector field of an observer moving at the curve $x^{i}(\tau,\lambda
=\lambda_{0}=$ const.$)$ interpreted as his world line, the vector field $\xi$
(and $\overline{\xi}$ respectively) could be chosen to lie in the subspace
orthogonal to $u$, i.e. $u$ and $\overline{\xi}$ could obey the condition
$g(u,\overline{\xi})=0$ and, therefore, $g(u,\xi)=0$, $\xi=\xi_{\perp
}=\overline{g}[h_{u}(\xi)]$, and\thinspace\thinspace\thinspace\thinspace
$\overline{\xi}=\overline{\xi}_{\perp}$.

4. In the next step, we could consider the vector field $\overline{\xi}$ as a
unit vector field in direction of the vector field $\xi$, i.e.
\begin{align*}
\overline{\xi}_{\perp}  & =n_{\perp}=\frac{\xi_{\perp}}{l_{\xi_{\perp}}%
}\,\,\,\,\,\,\text{,\thinspace\thinspace\thinspace\thinspace\thinspace
\thinspace\thinspace\thinspace\thinspace\thinspace\thinspace\thinspace
\thinspace\thinspace}g(u,n_{\perp})=0\,\,\,\,\,\,\text{,}\\
g(\overline{\xi}_{\perp},\overline{\xi}_{\perp})  & =g(n_{\perp},n_{\perp
})=\frac1{l_{\xi_{\perp}}^{2}}\cdot g(\xi_{\perp},\xi_{\perp})=\mp
\frac1{l_{\xi_{\perp}}^{2}}\cdot l_{\xi_{\perp}}^{2}=\mp1\text{ \thinspace
\thinspace,}\\
g(\overline{\xi}_{\perp},\overline{\xi}_{\perp})  & =g(\frac{\partial\lambda
}{\partial r}\cdot\xi_{\perp},\frac{\partial\lambda}{\partial r}\cdot
\xi_{\perp})=(\frac{\partial\lambda}{\partial r})^{2}\cdot g(\xi_{\perp}%
,\xi_{\perp})=\mp(\frac{\partial\lambda}{\partial r})^{2}\cdot l_{\xi_{\perp}%
}^{2}=\\
& =\mp1\,\,\,\,\,\text{,}%
\end{align*}
\[
(\frac{\partial\lambda}{\partial r})^{2}\cdot l_{\xi_{\perp}}^{2}%
=1\,\,\,\,\,\,\,\,\text{,\thinspace\thinspace\thinspace\thinspace
\thinspace\thinspace}l_{\xi_{\perp}}^{2}=\text{\thinspace\thinspace\thinspace
}(\frac{\partial\lambda}{\partial r})^{-2}\,\,\,\,\,\,\text{,\thinspace
\thinspace\thinspace\thinspace\thinspace\thinspace\thinspace\thinspace
\thinspace\thinspace}l_{\xi_{\perp}}=\pm\text{\thinspace\thinspace
\thinspace\thinspace\thinspace}(\frac{\partial\lambda}{\partial r}%
)^{-1}\,\,\,\,\,\text{.}
\]

After all above considerations for $\xi_{\perp}$ and $\overline{\xi}_{\perp}$
we obtain the relations
\begin{align*}
g(\overline{u},u)  & =g(u,u)\,\,\,\,\,\,\text{,}\\
g(\overline{u},\overline{\xi}_{\perp})  & =l_{v}\cdot g(\overline{\xi}_{\perp
},\overline{\xi}_{\perp})=l_{v}\cdot g(n_{\perp},n_{\perp})=\mp l_{v}%
\,\,\,\,\text{,}\\
g(\overline{u},\overline{u})  & =g(u,u)+l_{v}^{2}\cdot g(n_{\perp},n_{\perp
})=\\
& =\pm l_{u}^{2}\mp l_{v}^{2}=\frac{ds^{2}}{d\tau^{2}}\,\,\,\,\text{,}%
\end{align*}
\[
\frac{ds^{2}}{d\tau^{2}}=g(\frac d{d\tau},\frac d{d\tau})=\pm l_{u}^{2}\mp
l_{v}^{2}=\pm l_{u}^{2}\cdot(1-\frac{l_{v}^{2}}{l_{u}^{2}})\,\,\,\,\,\text{.}
\]

Moreover,
\begin{align*}
dx^{i}  & =(u^{i}+l_{v}\cdot n_{\perp}^{i})\cdot d\tau+\frac{\partial
r}{\partial\lambda}\cdot d\lambda\cdot n_{\perp}^{i}\,\,\,\,\,\text{,}\\
d  & =d\tau\cdot(u+l_{v}\cdot n_{\perp})+\frac{\partial r}{\partial\lambda
}\cdot d\lambda\cdot n_{\perp}=\\
& =d\tau\cdot u+(d\tau\cdot l_{v}+\frac{\partial r}{\partial\lambda}\cdot
d\lambda)\cdot n_{\perp}=\\
& =d\tau\cdot u+dr\cdot n_{\perp}\,\,\,\,\text{,}\\
dr  & =d\tau\cdot l_{v}+\frac{\partial r}{\partial\lambda}\cdot d\lambda
\,\,\,\,\,\text{,}\\
\frac{dr(\tau,\lambda)}{d\tau}  & =l_{v}\,\,\,\,\,\text{,\thinspace
\thinspace\thinspace\thinspace\thinspace\thinspace\thinspace\thinspace
\thinspace}\frac{d\lambda}{d\tau}=0\,\,\,\,\,\text{,}\\
\overline{u}  & =u+l_{v}\cdot n_{\perp}\,\,\,\,\,\,\text{,\thinspace
\thinspace\thinspace\thinspace\thinspace\thinspace\thinspace\thinspace
\thinspace}g(\overline{u},u)=g(u,u)\,\,\,\,\text{,}\\
n_{\perp}  & =\overline{g}[h_{u}(n_{\perp})]\,\,\,\,\,\,\,\,\text{,}%
\end{align*}
\begin{align*}
ds^{2}  & =g(d,d)=d\tau^{2}\cdot g(u,u)+(d\tau\cdot l_{v}+\frac{\partial
r}{\partial\lambda}\cdot d\lambda)^{2}\cdot g(n_{\perp},n_{\perp})=\\
& =\pm d\tau^{2}\cdot l_{u}^{2}\mp(d\tau\cdot l_{v}+\frac{\partial r}%
{\partial\lambda}\cdot d\lambda)^{2}=\pm d\tau^{2}\cdot l_{u}^{2}\mp d\tau
^{2}\cdot l_{v}^{2}=\\
& =\pm d\tau^{2}\cdot l_{u}^{2}\mp dr^{2}==\pm(l_{u}^{2}\cdot d\tau^{2}%
-dr^{2})=\pm d\tau^{2}\cdot(l_{u}^{2}-\frac{dr^{2}}{d\tau^{2}})=\\
& =\pm d\tau^{2}\cdot(l_{u}^{2}-l_{v}^{2})\,\,\,\,\,\text{,}%
\end{align*}

Therefore,
\begin{align}
ds^{2}  & =g(d,d)=\pm d\tau^{2}\cdot(l_{u}^{2}-l_{v}^{2}%
)\,\,\,\,\,\,\,\text{,}\label{2.3}\\
\frac{ds^{2}}{d\tau^{2}}  & =g(\frac d{d\tau},\frac d{d\tau})=g(\overline
{u},\overline{u})=\pm l_{u}^{2}\cdot(1-\frac{l_{v}^{2}}{l_{u}^{2}%
})\,\,\,\,\,\text{.}\label{2.4}%
\end{align}

5. In the non-relativistic field theories the distance between two points
$P\in M$ and $\overline{P}\in M$ is defined as
\[
ds^{2}=\mp dr^{2}\,\,\,\,\,\,
\]
and $l_{u}=0$. This means that the distance between two neighboring points $P$
and $\overline{P}$ is the space distance measured between them in the rest
(proper) reference frame of the observer (with absolute value $l_{u}$ of his
velocity $u$ equal to zero). The time parameter $\tau$ is not considered as a
co-ordinate in space-time, but as a parameter, independent of the frame of
reference of the observer.

6. In the relativistic field theories and especially in the Einstein theory of
gravitation $dr$ is considered as the space distance between two neighboring
points $P$ and $\overline{P}$ and $l_{u}\cdot d\tau$ is interpreted as the
distance covered by a light signal in a time interval $d\tau$, measured by an
observer in his proper frame of reference (when the observer in it is at
rest). The quantity $l_{u}$ is usually interpreted as the absolute value $c$
of a light signal in vacuum, i.e. $l_{u}=c$, or $l_{u}$ is normalized to $1$,
i.e. $l_{u}=1$, if the proper time interval $d\tau$ is replaced with the
proper distance interval $ds=c\cdot d\tau$, i.e. $\overline{u}=\frac d{d\tau}$
is replaced with $\overline{u}^{\prime}=\frac d{c\cdot d\tau}=\frac d{ds}$,
\thinspace\thinspace\thinspace$ds=c\cdot d\tau$.

Therefore, there is a difference between the interpretation of the absolute
value $l_{u}$ of the velocity of an observer in classical and relativistic physics

(a) In classical physics, from the above consideration, it follows that
$l_{u}=0$ (observer at rest) and $ds=dr$ is the distance as space distance.
$l_{v} $ is the absolute value of the velocity between the observer at rest
and a point $\overline{P}$ in its neighborhood..

(b) In relativistic physics $l_{u}=c$, or $l_{u}=1$, and $l_{u}$ is not the
absolute value of the velocity of the observer but the velocity of the light
propagation which the observer could measure in his proper frame of reference.
If we wish to interpret $l_{u}$ as the absolute value of the velocity of the
observer himself we should assume that $l_{u}\neq c$ or $1$ (if the observer
is not moving with the speed of light).

\begin{itemize}
\item There is the possibility to identify $l_{u}$ with $l_{v}$ as the
absolute value of the velocity of the observer at a point $P$ at his world
line, measured with respect to a neighboring point $\overline{P}$ with the
same proper time as the point $P$. Under this assumption, the ordinary
differential becomes a null (isotropic) vector field [$g(d,d)=0$, $l_{d}=0$,
$l_{u}=l_{v}\neq0$] in the proper frame of reference of the observer.

\item We could also interpret $l_{u}$ as the absolute value of the velocity of
the observer with respect to another frame of reference or

\item we can consider $l_{u}$ as the absolute value of the velocity of a
signal coming to the observer with velocity, different from the velocity of
light. On the basis of the last assumption we can describe the propagation of
signals with propagation velocity different from the velocity of light (for
instance, the propagation of sound signals or (may be) gravitational signals).
\end{itemize}

If $l_{v}=0$ then $\overline{u}=u$ and $u$ could be

\begin{itemize}
\item the velocity vector field $u=\frac d{d\tau}$ of an observer ($l_{u}%
\neq0$, $u=l_{u}\cdot n_{\parallel}$) in his proper frame of reference along
his world line. Since in his proper frame of reference the observer is at
rest, $u$ could be interpreted as the velocity of a clock measuring the length
(proper time) of the world line by the use of the parameter $\tau$ or

\item the velocity of a signal detected or emitted by the observer.
\end{itemize}

\subsection{Measuring a distance in $(\overline{L}_{n},g)$-spaces}

A. If the notion of distance is introduced in a space-time modeled by a
$(\overline{L}_{n},g)$-space we have to decide \textit{what is the meaning of
the vector field }$u$\textit{\ as tangent vector to a trajectory interpreted
as the world line of an observer}. On the basis of the above consideration, we
have four possible answers for the meaning of the vector field $u$ as

1. Velocity vector field of a propagating signal in space-time identified with
the tangent vector field $u$ at the world line of an observer. The signal is
detected or emitted by the observer on his world line and the absolute value
$l_{u}$ of $u$ is identified with the absolute value of the velocity of the
signal in- or outcouming to the observer.

2. Velocity vector field of an observer moving in space-time. In this case
$l_{u}\neq0$ and the space-time should have a definite metric, i.e.
$Sgn\,g=\pm n$, $dimM=n$ (for instance, motion of an observer in an Euclidean
space considered as a model of space-time). The observer, moving in space-time
could consider processes happened in its subspace orthogonal to his velocity.
The observer will move in a flow and consider the characteristics of the flow
from his own frame of reference.

3. Velocity of a clock moving in space-time and determining the proper time in
the frame of reference of an observer. The velocity $u$ of the clock in
space-time is with fixed absolute value $l_{u}$, i.e. $l_{u}=$ const. The time
interval $d\tau$ measured by the clock corresponds to the length $ds$ of its
world line, i.e. $d\tau^{2}=\pm\,$const.$\cdot ds^{2}$. Under the assumption
for the constant velocity of the clock we consider in it a periodical process
which indicates the time interval $d\tau$ in the proper frame of reference of
the clock and of the observer respectively.

4. Velocity $u$ of a $(n-1)$-dimensional subspace moving in time with
$l_{u}\neq0$. If the subspace deform in some way, the deformations reflect on
the kinematic characteristics of the vector field $u$ and $u$ is used as an
indicator for the changing properties of the subspace, considered as the space
of an observer (laboratory) where a physical system is investigated. This type
of interpretation requires not only the existence of the velocity vector field
$u$ with $l_{u}\neq0$ but also the existence of (at least one) orthogonal to
$u$ vector field $\xi_{\perp}$, $g(u,\xi_{\perp})=0$, lying in the orthogonal
to $u$ subspace $T^{\perp u}(M)$.

All indicated interpretations could be used in solving different physical
problems related to motions of physical systems in space-time.

B. After introducing the notion of distance, the question arises \textit{how a
space distance between two points in a space could be measured}. We could
distinguish three types of measurements:

1. Direct measurements by using a measuring device (e.g. a roulette, a linear
(running) meter, yard-measure-stick etc.)

2. Direct measurements by sending signals from a basic point to a fixed point
of space and detecting at the basic point the reflected by the fixed point signal.

3. Indirect measurement by receiving signals from a fixed point of space
without sending a signal to it.

Let us now consider every type of measurements more closely.

1. \textit{Direct measurements by using a measuring device.} The space
distance between two points $A$ and $B$ in a space could be measured by a
second observer moving from point $A$ (where the first observer is at rest) to
point $B$ in space. At the same time, the second observer moves in time from
point $B$ to point $B^{\prime}$. The space distance measured by the observer
with world line $AA^{\prime}$ could be denoted as $\vartriangle r=AB$ and the
time period passed as $\vartriangle\tau=AA^{\prime}$. This is a direct
measurement of the space distance $AB=\vartriangle r$ from point $A$ to point
$B$ in the space during the time $AA^{\prime}=\vartriangle\tau$. It is
\textit{assumed} that point $A$ and point $B$ are at rest during the
measurement. Instead of measuring the space distance $AB$ the observers
measure the space distance $A^{\prime}B^{\prime}$ which exists at the time
$\tau+\vartriangle\tau$ if the measurement has began at the time $\tau$ from
the point of the first observer with world line $AA^{\prime}$.

2. \textit{Direct measurements by sending signals from a basic point to a
fixed point of space and detecting at the basic point the reflected by the
fixed point signal.} The space distance between two points $A$ and $B$ in a
space could be measured by a sending a signal with velocity with absolute
value $l_{u}\neq0$. Then $AB$ of the curve $x^{i}(\tau=\tau_{0},r)$ through
point $B$ is the distance $\vartriangle r$ at the time $\tau(A)=\tau_{0}$ and
$\tau(B)=\tau_{0}$.

$A^{\prime}B^{\prime}$ of the curve $x^{i}(\tau=\tau_{0}+\vartriangle\tau,r)$
is the space distance $\vartriangle r^{\prime}$ at the time $\tau(A^{\prime
})=\tau_{1}$. At this time the signal is received at point $B^{\prime}$ which
is point $B$ at the time $\tau_{1}$, i.e. $\tau(B^{\prime})=\tau_{1}$.
$B^{\prime}A^{\prime}$ is the space distance between $B$ and $A$ at the time
$\tau(A^{\prime})=\tau_{1}$, where $\tau(B^{\prime})=\tau_{1}$, $\tau
(B^{\prime\prime})=\tau_{2}$. At the time $(\tau_{2})$ the point $B(\tau_{0})$
will be moved in the time to point $B^{\prime\prime}(\tau_{2})$. The signal
will be propagated

(a) for the time interval $AA^{\prime}=\tau_{1}-\tau_{0}$ to the point
$B^{\prime}$ at the time $\tau_{1}$ at the space distance $\vartriangle
r=l_{u}\cdot(\tau_{1}-\tau_{0})$, where $l_{u}$ is the velocity of the signal
measured by the observer with world line $AA^{\prime}$.

(b) for the time interval $A^{\prime}A^{\prime\prime}$ from point $B^{\prime}$
at the time $\tau_{1}$ to the point $A^{\prime\prime}$ at the time $\tau_{2}$
at a space distance $l_{u}\cdot(\tau_{2}-\tau_{1})$. The whole space distance
covered by the signal in the time interval $AA^{\prime}A^{\prime\prime
}=\vartriangle\tau=\tau_{2}-\tau_{0}$ is $l_{u}\cdot(\tau_{2}-\tau_{0})=$
$l_{u}\cdot(\tau_{2}-\tau_{1})+l_{u}\cdot(\tau_{1}-\tau_{0})$.

If we now \textit{assume} that point $A$ and point $B$ are at rest to each
other and the space distance between them does not change in the time then
\[
l_{u}\cdot(\tau_{2}-\tau_{1})=l_{u}\cdot(\tau_{1}-\tau_{0})
\]
and
\[
l_{u}\cdot(\tau_{2}-\tau_{0})=2\cdot l_{u}\cdot(\tau_{1}-\tau_{0})=2\cdot
A^{\prime}B^{\prime}(\tau_{1})=2\cdot AB(\tau_{0})\,\,\,\text{.}
\]

Therefore, the space distance between point $A$ and point $B$ (at any time, if
both the points are at rest to each other) is
\[
AB=\frac12\cdot l_{u}\cdot(\tau_{2}-\tau_{0})\,\,\,\,\text{,}
\]
where $\vartriangle\tau=\tau_{2}-\tau_{0}$ is the time interval for the
propagation of a signal from point $A$ to point $B$ and from point $B$ back to
point $A$.

3. \textit{Indirect measurement by receiving signals from a fixed point of
space without sending a signal to it}. If the space distance between point $A
$ and point $B$ is changing in the time and at point $B$ there is an emitter
then the frequency of the emitter will change in the time related to the
centrifugal (centripetal) or Coriolis' velocity between both the points $A$
and $B$. Therefore, a criteria for no relative motion between two (space)
points (points with one and the same proper time) could be the lack of change
of the frequency of the signals emitted from the second point $B$ to the basic
point $A$. [But there could be motions of an emitter which could so change its
frequency that the changes compensate each other and the observer at the basic
point $A$ could come to the conclusion that there is no motions between points
$A$ and $B$.]

If an emitter at point $B(\tau_{0})$ emits a signal with velocity
$\overline{u}$ and frequency $\overline{\omega}$ then this signal will be
received (detected) at the point $A^{\prime}(\tau_{1})$ after a time interval
$AA^{\prime}=\vartriangle\tau=\tau_{1}-\tau_{0}$ by an observer (detector)
moving in the time interval $\vartriangle\tau$ from point $A(\tau_{0})$ to
point $A^{\prime}(\tau_{1})$ on his world line $x^{i}(\tau)$. If the emitter
is moving relatively to point $A$ with relative velocity $_{rel}v$ the
detected at the point $A^{\prime}$ frequency $\omega$ will differ from the
emitted frequency $\overline{\omega}$. If both the points $A$ and $B$ are at
rest to each other then $\overline{\omega}=\omega$.

C. The question arises \textit{how can we find the space distance between two
points }$A$\textit{\ and }$B$\textit{\ lying in such a way in the space that
only signals emitted from the one point (point }$B$\textit{) could be detected
at the basic point (point }$A$\textit{), where an observer detects the signal
from point }$B$\textit{.} First of all, if we knew the propagation velocity
$l_{u}$ of a signal and the difference $\overline{\omega}-\omega$ between the
emitted frequency $\overline{\omega}$ and the detected frequency $\omega$ we
can try to find out the relative velocity between the emitter (at a point $B$)
and the observer (at a point $A$). For doing that we will need relations
between the difference $\overline{\omega}-\omega$ and the relative velocity
between both the points. Such relations could be found on the basis of the
structures of the relative velocity and its decomposition in centrifugal
(centripetal) relative velocity and Coriolis' relative velocity.

\section{Kinematic effects related to the relative velocity}

1. Let us now consider the change of a null vector field $\widetilde{k}$ under
the influence of the relative velocity of the corresponding emitter and its
frequency with respect to an observer detecting the emitted radiation by the emitter.

Let $\overline{k}_{\perp}$ be the orthogonal to $u$ part of the null vector
field $\overline{k}$ corresponding to the null vector field $\widetilde{k}$
after the influence of the relative velocity $_{rel}v$%
\begin{align}
\overline{k}  & =\widetilde{k}+\,_{rel}k\text{ \thinspace,\thinspace
\thinspace\thinspace\thinspace\thinspace}\overline{k}=\overline{k}_{\parallel
}+\overline{k}_{\perp}\,\,\,\,\text{,\thinspace\thinspace\thinspace
\thinspace\thinspace}\widetilde{k}=k_{\parallel}+k_{\perp}%
\,\,\,\,\text{,\thinspace\thinspace\thinspace\thinspace\thinspace}%
_{rel}k=\,_{rel}k_{\parallel}+\,_{rel}k_{\perp}\text{ ,}\label{3.1}\\
\overline{k}_{\perp}  & =\widetilde{k}_{\perp}+\,_{rel}k_{\perp}%
\text{\thinspace\thinspace\thinspace,}\label{3.2}%
\end{align}
where $_{rel}k$ depends on the relative velocity $_{rel}v$.

If $\overline{k}=\widetilde{k}+\,_{rel}k$ then
\begin{align*}
g(\overline{k},\overline{k})  & =0=g(_{rel}k,\,_{rel}k)+2\cdot g(\widetilde
{k},\,_{rel}k)\,\,\,\,\,\text{,}\\
g(_{rel}k,\,_{rel}k)  & =-2\cdot g(\widetilde{k},\,_{rel}k)\,\,\,\,\,\text{.}%
\end{align*}

If $_{rel}k=C\cdot\widetilde{k}$ then $\overline{k}=\widetilde{k}%
+\,C\cdot\widetilde{k}=(1+C)\cdot\widetilde{k}$ and $g(\overline{k}%
,\overline{k})=0=(1+C)^{2}\cdot g(\widetilde{k},\widetilde{k})=0$ for $\forall
C\in C^{r}(M)$. Therefore, the assumption that $_{rel}k=C\cdot\widetilde{k}$,
$C\neq0$, leads to a mapping of the vector field $\widetilde{k}$ as a null
vector field into a new null vector field $\overline{k}$ under the influence
of the relative velocity between emitter and detector. Then
\begin{align}
_{rel}k  & =C\cdot\widetilde{k}=C\cdot(k_{\parallel}+k_{\perp}%
)\text{\thinspace\thinspace\thinspace\thinspace\thinspace,}\label{3.3}\\
\vartriangle\omega & =g(u,\,_{rel}k)=C\cdot g(u,\widetilde{k})=C\cdot
g(u,k_{\parallel})=C\cdot\omega\text{ \thinspace\thinspace\thinspace
,}\label{3.4}\\
g(\overline{k},u)  & =g(\widetilde{k},u)+g(_{rel}k,u)\text{ ,}\label{3.5}\\
\overline{\omega}  & =\omega+C\cdot\omega=(1+C)\cdot\omega\,\,\,\,\,\text{,}%
\label{3.6}\\
\vartriangle\omega & =C\cdot\omega=\overline{\omega}-\omega\text{
\thinspace\thinspace,}\label{3.7}%
\end{align}
\begin{align}
g(_{rel}k,\widetilde{n}_{\perp})  & =g(C\cdot\widetilde{k},\widetilde
{n}_{\perp})=g(_{rel}k_{\perp},\widetilde{n}_{\perp})=C\cdot g(k_{\perp
},\widetilde{n}_{\perp})=\label{3.9}\\
& =C\cdot\frac\omega{l_{u}}\text{\thinspace\thinspace\thinspace\thinspace
\thinspace,}\label{3.10}%
\end{align}
\begin{align}
_{rel}k_{\parallel}  & =\pm\frac{\vartriangle\omega}{l_{u}}\cdot n_{\parallel
}=\pm C\cdot\frac\omega{l_{u}}\cdot n_{\parallel}%
\,\,\,\,\,\,\,\text{,\thinspace\thinspace}\label{3.11}\\
\text{\thinspace\thinspace\thinspace\thinspace\thinspace\thinspace\thinspace
}_{rel}k_{\perp}  & =\pm\frac{\vartriangle\omega}{l_{u}}\cdot\widetilde
{n}_{\perp}=\pm C\cdot\frac\omega{l_{u}}\cdot\widetilde{n}_{\perp
}\,\,\,\,\,\text{.}\label{3.12}%
\end{align}

The problem arises how can we find the invariant function (factor) $C$ before
$\widetilde{k}$ depending on the relative velocity $_{rel}v$ and the velocity
of the signal $l_{u}$. Since $l_{_{rel}k_{\parallel}}=l_{_{rel}k_{\perp}}$ we
can find in this way the whole structure of $_{rel}k=\,_{rel}k_{\parallel
}+\,_{rel}k_{\perp}$.

In the previous sections we have considered the representation of
$\widetilde{k}$ as $\widetilde{k}=k_{\parallel}+k_{\perp}$, where
\begin{align*}
k_{\parallel}  & =\pm l_{k_{\parallel}}\cdot n_{\parallel}=\pm\frac
\omega{l_{u}}\cdot n_{\parallel}\,\,\,\,\text{,\thinspace\thinspace
\thinspace\thinspace\thinspace\thinspace\thinspace\thinspace\thinspace
\thinspace\thinspace\thinspace\thinspace\thinspace\thinspace\thinspace
}l_{k_{\parallel}}=\frac\omega{l_{u}}\,\,\,\,\text{,\thinspace\thinspace
\thinspace\thinspace\thinspace\thinspace\thinspace}\\
\text{\thinspace}k_{\perp}  & =\mp l_{k_{\perp}}\cdot\widetilde{n}_{\perp}%
=\mp\frac\omega{l_{u}}\,\cdot\widetilde{n}_{\perp}\,\,\,\text{,\thinspace
\thinspace\thinspace\thinspace\thinspace}l_{k_{\perp}}=\frac\omega{l_{u}%
}=\text{\thinspace}l_{k_{\parallel}}\,\,\,\,\text{.}%
\end{align*}

The unit vector $\widetilde{n}_{\perp}$ is orthogonal to the vector $u$, i.e.
$g(u,\widetilde{n}_{\perp})=0$ because of $g(u,k_{\perp})=\mp l_{k_{\perp}%
}\cdot g(u,\widetilde{n}_{\perp})=0$,\thinspace\thinspace\thinspace
$l_{k_{\perp}}\neq0$, $l_{u}\neq0$. Therefore, $g(\widetilde{n}_{\perp
},\widetilde{n}_{\perp})=\mp1=\mp l_{\widetilde{n}_{\perp}}^{2}$,
$l_{\widetilde{n}_{\perp}}\neq0$.

We can represent the unit vector $\widetilde{n}_{\perp}$ (orthogonal to $u$)
in two parts: one part collinear to the vector field $\xi_{\perp}$ (orthogonal
to $u$) and one part orthogonal to the vectors $u$ and $\xi_{\perp}$, i.e.
\begin{align}
\widetilde{n}_{\perp}  & =\alpha\cdot n_{\perp}+\beta\cdot m_{\perp
}\,\,\,\,\,\,\,\text{,\thinspace\thinspace\thinspace\thinspace\thinspace
\thinspace\thinspace\thinspace\thinspace\thinspace}\label{3.13}\\
g(\widetilde{n}_{\perp},\widetilde{n}_{\perp})  & =\mp1=\mp l_{\widetilde
{n}_{\perp}}^{2}\,\,\,\text{,\thinspace\thinspace\thinspace\thinspace
\thinspace\thinspace}l_{\widetilde{n}_{\perp}}>0\,\,\,\,\,\,\text{,\thinspace
\thinspace\thinspace\thinspace\thinspace\thinspace\thinspace\thinspace
\thinspace}l_{\widetilde{n}_{\perp}}=1\,\,\,\,\text{,}\label{3.14}%
\end{align}
where
\begin{align}
n_{\perp}  & =\frac{\xi_{\perp}}{l_{\xi_{\perp}}}\,\,\,\,\,\text{,\thinspace
\thinspace\thinspace\thinspace\thinspace\thinspace\thinspace}g(n_{\perp
},u)=0\,\,\,\,\,\text{,\thinspace\thinspace\thinspace\thinspace\thinspace
}g(n_{\perp},n_{\perp})=\mp1=\mp l_{n_{\perp}}^{2}\,\,\text{,}\label{3.15}\\
l_{n_{\perp}}  & >0\,\,\,\,\text{,\thinspace\thinspace\thinspace
\thinspace\thinspace\thinspace}l_{n_{\perp}}=1\label{3.16}\\
m_{\perp}  & =\frac{v_{c}}{l_{v_{c}}}\,\,\,\,\,\text{,\thinspace
\thinspace\thinspace\thinspace\thinspace\thinspace\thinspace}g(m_{\perp
},u)=0\,\,\,\,\,\text{,\thinspace\thinspace\thinspace\thinspace\thinspace
\thinspace}g(m_{\perp},\xi_{\perp})=0\,\,\text{.}\label{3.17}%
\end{align}

The vector field $v_{c}$ is the Coriolis velocity vector field orthogonal to
$u $ and to the centrifugal (centripetal) velocity $v_{z}$ collinear to
$\xi_{\perp}$. Since
\begin{equation}
v_{c}=\pm l_{v_{c}}\cdot m_{\perp}\text{\thinspace\thinspace\thinspace
\thinspace\thinspace\thinspace\thinspace,\thinspace\thinspace\thinspace
\thinspace\thinspace\thinspace\thinspace\thinspace\thinspace\thinspace}%
g(v_{c},v_{c})=\mp l_{v_{c}}^{2}\text{\ \ \ \ \ ,}\label{3.18}%
\end{equation}
we have
\begin{equation}
g(m_{\perp},m_{\perp})=\mp1=\mp l_{m_{\perp}}^{2}\,\,\,\,\,\text{,\thinspace
\thinspace\thinspace\thinspace\thinspace\thinspace\thinspace\thinspace
}l_{m_{\perp}}>0\,\,\,\,\,\,\,\text{,\thinspace\thinspace\thinspace
\thinspace\thinspace\thinspace}l_{m_{\perp}}=1\text{ \thinspace\thinspace
\thinspace.}\label{3.19}%
\end{equation}

The Coriolis velocity $v_{c}$ is related to the change of the vector
$\xi_{\perp}$ in direction orthogonal to $u$ and $\xi_{\perp}$.

Since $\widetilde{n}_{\perp}$ is a unit vector as well as the vectors
$n_{\perp}$ and $m_{\perp}$, and, further, $g(n_{\perp},m_{\perp})=0$, we
obtain
\begin{align*}
g(\widetilde{n}_{\perp},\widetilde{n}_{\perp})  & =\mp1=g(\alpha\cdot
n_{\perp}+\beta\cdot m_{\perp},\alpha\cdot n_{\perp}+\beta\cdot m_{\perp})=\\
& =\alpha^{2}\cdot g(n_{\perp},n_{\perp})+\beta^{2}\cdot g(m_{\perp},m_{\perp
})=\\
& =\mp\alpha^{2}\mp\beta^{2}\,\,\,\text{.}%
\end{align*}

Therefore,
\begin{equation}
\alpha^{2}+\beta^{2}=1\,\,\,\,\text{.}\label{3.20}%
\end{equation}

On the other side,
\begin{align*}
g(\widetilde{n}_{\perp},n_{\perp})  & =g(\alpha\cdot n_{\perp}+\beta\cdot
m_{\perp},n_{\perp})=\\
& =\alpha\cdot g(n_{\perp},n_{\perp})=\mp\alpha\,\,\,\,\,\text{,}\\
g(\widetilde{n}_{\perp},m_{\perp})  & =g(\alpha\cdot n_{\perp}+\beta\cdot
m_{\perp},m_{\perp})=\\
& ==\beta\cdot g(n_{\perp},n_{\perp})=\mp\beta\,\,\,\text{.}%
\end{align*}
i.e.
\begin{align}
\alpha & =\mp g(\widetilde{n}_{\perp},n_{\perp})=\mp l_{\widetilde{n}_{\perp}%
}\cdot l_{n_{\perp}}\cdot cos(\widetilde{n}_{\perp},n_{\perp})=\mp
\,cos(\widetilde{n}_{\perp},n_{\perp})\,\,\,\,\text{,}\label{3.21}\\
\beta & =\mp g(\widetilde{n}_{\perp},m_{\perp})=\mp l_{\widetilde{n}_{\perp}%
}\cdot l_{m_{\perp}}\cdot cos(\widetilde{n}_{\perp},m_{\perp})=\mp
cos(\widetilde{n}_{\perp},m_{\perp})\,\,\,\,\text{.}\label{3.22}%
\end{align}

Therefore, $\alpha$ and $\beta$ appear as direction cosines of $n_{\perp}$ and
$m_{\perp}$ with respect to the unit vector $\widetilde{n}_{\perp}$. Since
\[
cos^{2}(\widetilde{n}_{\perp},n_{\perp})+cos^{2}(\widetilde{n}_{\perp
},m_{\perp})=1\,\,\,\,\,\text{,}
\]
it follows that
\begin{align}
cos^{2}(\widetilde{n}_{\perp},m_{\perp})  & =1-cos^{2}(\widetilde{n}_{\perp
},n_{\perp})=1-sin^{2}(\widetilde{n}_{\perp},m_{\perp})=sin^{2}\,(\widetilde
{n}_{\perp},n_{\perp})\,\,\text{,}\nonumber\\
sin^{2}(\widetilde{n}_{\perp},m_{\perp})  & =cos^{2}(\widetilde{n}_{\perp
},n_{\perp})\,\,\,\,\,\text{,}\nonumber\\
cos(\widetilde{n}_{\perp},m_{\perp})  & =\pm\,sin\,(\widetilde{n}_{\perp
},n_{\perp})\,\,\,\,\text{,}\nonumber\\
\alpha & =\mp\,cos(\widetilde{n}_{\perp},n_{\perp})\,\,\,\,\text{,}%
\label{3.23}\\
\beta & =\mp sin\,(\widetilde{n}_{\perp},n_{\perp})\,\,\,\,\,\text{.}%
\label{3.24}%
\end{align}
\begin{align}
\widetilde{n}_{\perp}  & =\alpha\cdot n_{\perp}+\beta\cdot m_{\perp
}\,=\nonumber\\
& =\mp[cos(\widetilde{n}_{\perp},n_{\perp})\,\cdot n_{\perp}+sin\,(\widetilde
{n}_{\perp},n_{\perp})\,\cdot m_{\perp}]\,\,\,\text{.}\label{3.25}%
\end{align}

Further, since $k_{\perp}=\mp l_{k_{\perp}}\cdot\widetilde{n}_{\perp}$ then
(see above)
\begin{align*}
g(k_{\perp},k_{\perp})  & =l_{k_{\perp}}^{2}\cdot g(\widetilde{n}_{\perp
},\widetilde{n}_{\perp})=\mp l_{k_{\perp}}^{2}\,\,\,\,\,\,\text{,\thinspace
\thinspace\thinspace\thinspace\thinspace\thinspace}g(\widetilde{n}_{\perp
},\widetilde{n}_{\perp})=\mp1\,\,\,\,\,\text{,}\\
\text{\thinspace}k_{\perp}  & =\mp\frac\omega{l_{u}}\cdot\widetilde{n}_{\perp
}\,\,\,\,\,\,\text{,\thinspace\thinspace\thinspace\thinspace\thinspace
\thinspace\thinspace\thinspace\thinspace\thinspace\thinspace\thinspace
\thinspace\thinspace}k_{\parallel}=\pm\frac\omega{l_{u}}\cdot n_{\parallel
}\,\,\,\,\,\,\,\text{,\thinspace\thinspace\thinspace\thinspace\thinspace
\thinspace\thinspace\thinspace\thinspace\thinspace\thinspace}l_{k_{\perp}%
}=\frac\omega{l_{u}}=l_{k_{\parallel}}\,\,\,\,\text{,}\\
g(\widetilde{n}_{\perp},k_{\perp})  & =\mp l_{k_{\perp}}\cdot g(\widetilde
{n}_{\perp},\widetilde{n}_{\perp})=\frac\omega{l_{u}}\,\,=l_{k_{\parallel}%
}=l_{k_{\perp}}\,\,\,\,\,\text{.}%
\end{align*}

2. For the contravariant null vector field $\overline{k}$ we have analogous
relations as for the contravariant null vector field $\widetilde{k}$ (just
changing $\widetilde{k}$ with $\overline{k}$ and $\omega$ with $\overline
{\omega}$)
\begin{equation}
\overline{k}=\overline{k}_{\parallel}+\overline{k}_{\perp}%
\,\,\,\,\,\,\,\text{,\thinspace\thinspace\thinspace\thinspace\thinspace
\thinspace\thinspace\thinspace\thinspace\thinspace\thinspace\thinspace
}\overline{\omega}=g(u,\overline{k})\,\,\,\,\text{,}\label{3.26}%
\end{equation}
\begin{align}
\overline{k}_{\parallel}  & =\pm\frac{\overline{\omega}}{l_{u}}\cdot
n_{\parallel}\,\,\,\,\,\,\,\,\,\,\,\text{,\thinspace\thinspace\thinspace
\thinspace\thinspace\thinspace\thinspace\thinspace\thinspace\thinspace
\thinspace\thinspace}l_{\overline{k}_{\parallel}}=\frac{\overline{\omega}%
}{l_{u}}\,\,\,\,\text{,}\label{3.27}\\
\overline{k}_{\perp}  & =\mp\frac{\overline{\omega}}{l_{u}}\cdot\widetilde
{n}_{\perp}\,\,\,\,\,\text{,\thinspace\thinspace\thinspace\thinspace
\thinspace\thinspace\thinspace\thinspace\thinspace\thinspace}l_{\overline
{k}_{\perp}}=\frac{\overline{\omega}}{l_{u}}=l_{\overline{k}_{\parallel}%
}\,\,\,\,\text{,}\label{3.28}%
\end{align}
\begin{equation}
g(\widetilde{n}_{\perp},\overline{k}_{\perp})=\mp l_{\overline{k}_{\perp}%
}\cdot g(\widetilde{n}_{\perp},\widetilde{n}_{\perp})=\frac{\overline{\omega}%
}{l_{u}}\,\,=l_{\overline{k}_{\parallel}}=l_{\overline{k}_{\perp}%
}\,\,\,\,\,\,\,\,\text{.}\label{3.29}%
\end{equation}

From $\overline{k}=\widetilde{k}+\,_{rel}k$, $\overline{k}_{\perp}%
=\widetilde{k}_{\perp}+\,_{rel}k_{\perp}$, and
\begin{equation}
\mp\frac{\overline{\omega}}{l_{u}}\cdot\widetilde{n}_{\perp}=\mp\frac
\omega{l_{u}}\cdot\widetilde{n}_{\perp}+\,_{rel}k_{\perp}\,\,\,\,\text{,}%
\label{3.30}%
\end{equation}
it follows that
\begin{align}
g(\widetilde{n}_{\perp},\overline{k})  & =g(\widetilde{n}_{\perp
},k)+g(\widetilde{n}_{\perp},\,_{rel}k)\,\,\,\,\,\,\text{,}\nonumber\\
g(\widetilde{n}_{\perp},\overline{k}_{\perp})  & =g(\widetilde{n}_{\perp
},k_{\perp})+g(\widetilde{n}_{\perp},\,_{rel}k_{\perp})\,\,\,\,\text{,}%
\label{3.31}\\
\frac{\overline{\omega}}{l_{u}}  & =\frac\omega{l_{u}}+g(_{rel}k_{\perp
},\widetilde{n}_{\perp})\,\,\,\,\text{.}\label{3.32}%
\end{align}

The vectors $\overline{k}_{\perp}$ and $k_{\perp}$ are collinear to each
other. This means that the last term $g(_{rel}k_{\perp},\widetilde{n}_{\perp
})$ in the previous expression should be proportional to $k_{\perp}$, i.e.
$_{rel}k_{\perp}=C\cdot k_{\perp}$. At the same time, it should contain in its
factor $C$ before $k_{\perp}$a dimensionless term describing the influence of
the relative velocity $_{rel}v$ on the null vector field $\widetilde{k}$. This
term should take into account the fact that $_{rel}v$ should act in the
direction of $\widetilde{n}_{\perp}$ if its influence is on $k_{\perp}%
=\mp\frac\omega{l_{u}}\cdot\widetilde{n}_{\perp}$and as a result it leads to
$\overline{k}_{\perp}=\mp\frac{\overline{\omega}}{l_{u}}\cdot\widetilde
{n}_{\perp}$, i.e. it leads to a new vector $\overline{k}_{\perp}$ collinear
to $k_{\perp}$. On this basis, the conclusion could be made that we can define
$_{rel}k_{\perp}$ in the form
\begin{equation}
_{rel}k_{\perp}=\frac1{l_{u}}\cdot g(_{rel}v,\widetilde{n}_{\perp})\cdot
k_{\perp}\,\,\,\text{.}\label{3.33}%
\end{equation}

Therefore, from the expressions
\begin{align*}
\frac{\overline{\omega}}{l_{u}}  & =\frac\omega{l_{u}}+g(_{rel}k_{\perp
},\widetilde{n}_{\perp})\,=\frac\omega{l_{u}}+\frac1{l_{u}}\cdot
g(_{rel}v,\widetilde{n}_{\perp})\cdot g(k_{\perp},\widetilde{n}_{\perp})=\\
& =\frac\omega{l_{u}}+\frac1{l_{u}}\cdot g(_{rel}v,\widetilde{n}_{\perp}%
)\cdot\frac\omega{l_{u}}\,\,\,
\end{align*}
follows the relation%

\begin{equation}
\overline{\omega}=\omega+\frac1{l_{u}}\cdot g(_{rel}v,\widetilde{n}_{\perp
})\cdot\omega\,\,\,\,\text{.}\label{3.34}%
\end{equation}

3. Since $_{rel}v=v_{z}+v_{c}$ and $\widetilde{n}_{\perp}=\alpha\cdot
n_{\perp}+\beta\cdot m_{\perp}$, we can find the explicit form of the term
with $g(_{rel}v,\widetilde{n}_{\perp})$. By the use of the relations
\cite{Manoff-6}
\begin{align}
g(_{rel}v,\widetilde{n}_{\perp})  & =g(v_{z}+v_{c},\widetilde{n}_{\perp
})=g(v_{z},\widetilde{n}_{\perp})+g(v_{c},\widetilde{n}_{\perp})\,\,\,\text{,}%
\label{3.35}\\
g(v_{z},\widetilde{n}_{\perp})  & =g(v_{z},\alpha\cdot n_{\perp}+\beta\cdot
m_{\perp})=\alpha\cdot g(v_{z},n_{\perp})+\beta\cdot g(v_{z},m_{\perp
})=\nonumber\\
& =\alpha\cdot g(v_{z},n_{\perp})\,\,\,\,\text{,}\label{3.36}\\
g(v_{c},\widetilde{n}_{\perp})  & =g(v_{c},\alpha\cdot n_{\perp}+\beta\cdot
m_{\perp})=\alpha\cdot g(v_{c},n_{\perp})+\beta\cdot g(v_{c},m_{\perp
})=\nonumber\\
& =\beta\cdot g(v_{c},m_{\perp})\,\,\,\text{,}\label{3.37}%
\end{align}
\begin{align}
v_{z}  & =\pm l_{v_{z}}\cdot n_{\perp}=H\cdot l_{\xi_{\perp}}\cdot n_{\perp
}\,\,\,\,\,\,\text{,\thinspace\thinspace\thinspace\thinspace\thinspace
\thinspace\thinspace\thinspace\thinspace}n_{\perp}=\frac{\xi_{\perp}}%
{l_{\xi_{\perp}}}\,\,\,\,\text{,}\label{3.38}\\
v_{c}  & =\pm l_{v_{c}}\cdot m_{\perp}=H_{c}\cdot l_{\xi_{\perp}}\cdot
m_{\perp}\,\,\,\,\,\text{,\thinspace\thinspace\thinspace\thinspace
\thinspace\thinspace}m_{\perp}=\frac{v_{c}}{l_{v_{c}}}\,\,\,\text{.}%
\,\,\label{3.39}%
\end{align}

\textit{Remark}. The signs $\pm$ before $l_{v_{z}}$ and $l_{v_{c}}$ are not
related to the signature of the metric $g$. They are showing the direction of
$v_{z}$ and $v_{c}$ with respect to the units vectors $n_{\perp}$ and
$m_{\perp}$ respectively.

For $g(_{rel}v,\widetilde{n}_{\perp})$ we obtain
\begin{align}
g(_{rel}v,\widetilde{n}_{\perp})  & =g(v_{z}+v_{c},\alpha\cdot n_{\perp}%
+\beta\cdot m_{\perp})=\label{3.40}\\
& =\alpha\cdot g(v_{z},n_{\perp})+\beta\cdot g(v_{c},m_{\perp})=\nonumber\\
& =\alpha\cdot g(\pm l_{v_{z}}\cdot n_{\perp},n_{\perp})+\beta\cdot g(\pm
l_{v_{c}}\cdot m_{\perp},m_{\perp})=\nonumber\\
& =\pm\alpha\cdot l_{v_{z}}\cdot g(n_{\perp},n_{\perp})\pm\beta\cdot l_{v_{c}%
}\cdot g(m_{\perp},m_{\perp})=\nonumber\\
& =-\alpha\cdot l_{v_{z}}-\beta\cdot l_{v_{c}}=-(\alpha\cdot l_{v_{z}}%
+\beta\cdot l_{v_{c}})\,\,\,\,\text{.}\label{3.41}%
\end{align}

For $\overline{\omega}=\omega+\frac1{l_{u}}\cdot g(_{rel}v,\widetilde
{n}_{\perp})\cdot\omega$ the relation follows
\begin{equation}
\overline{\omega}=\omega-\frac1{l_{u}}\cdot(\alpha\cdot l_{v_{z}}+\beta\cdot
l_{v_{c}})\cdot\omega\,\,\,\,\text{.}\label{3.42}%
\end{equation}

Since $l_{v_{z}}=\pm H\cdot l_{\xi_{\perp}}$ and $l_{v_{c}}=\pm H_{c}\cdot
l_{\xi_{\perp}}$, it follows further
\begin{equation}
\overline{\omega}=\omega\mp\frac1{l_{u}}\cdot(\alpha\cdot H+\beta\cdot
H_{c})\cdot l_{\xi_{\perp}}\cdot\omega\,\,\,\,\,\,\,\,\text{.}\label{3.43}%
\end{equation}

\textit{Remark}. The explicit forms of $H$ and $H_{c}$ will be given below in
the considerations of the Hubble and the aberration effects.

Therefore, for $\overline{\omega}$ and $\omega$ we have the relations

(a)
\[
\overline{\omega}=\omega-\frac1{l_{u}}\cdot(\alpha\cdot l_{v_{z}}+\beta\cdot
l_{v_{c}})\cdot\omega\,\,\,\text{,}
\]

(b)
\[
\overline{\omega}=\omega\mp\frac1{l_{u}}\cdot(\alpha\cdot H+\beta\cdot
H_{c})\cdot l_{\xi_{\perp}}\cdot\omega\,\,\,\,\,\,\,\,\text{.}
\]

On the basis of the relations (a) and (b) different kinematic effects could be
considered related to the Doppler effect, to the Hubble effect, and to the
aberration effect.

\subsection{Standard (longitudinal) and transversal Doppler effects}

The expression for $\overline{\omega}$ could also be written in the form
\begin{equation}
\overline{\omega}=[1-(\alpha\cdot\frac{l_{v_{z}}}{l_{u}}+\beta\cdot
\frac{l_{v_{c}}}{l_{u}})]\cdot\omega\,\,\,\,\,\text{,}\label{3.44}%
\end{equation}
where $\overline{\omega}$ is the frequency of an emitter moving with
centrifugal (centripetal) velocity $v_{z}=\pm$ $l_{v_{z}}\cdot n_{\perp}$ and
with Coriolis' velocity $v_{c}=\pm l_{v_{c}}\cdot m_{\perp}$ relative to an
observer (with detector) \cite{Manoff-6}. The emitted signals propagate with
velocity $u=l_{u}\cdot n_{\parallel}$ with respect to the observer, where $u$
is the tangent vector to the world line of the observer (detector). The
detected signals are with frequency $\omega$.

(a) If $l_{v_{c}}=0$ and the emitter moves only away or to the observer, i.e.
if $\alpha=\pm1$, $\beta=0$, then
\begin{equation}
\overline{\omega}=(1\pm\frac{l_{v_{z}}}{l_{u}})\cdot\omega\,\,\,\,\text{.}%
\label{3.45}%
\end{equation}

For $\overline{\omega}>\omega$%
\begin{equation}
\overline{\omega}=(1+\frac{l_{v_{z}}}{l_{u}})\cdot\omega\,\,\,\,\,\,\text{.}%
\label{3.46}%
\end{equation}

For $\overline{\omega}<\omega$%
\begin{equation}
\overline{\omega}=(1-\frac{l_{v_{z}}}{l_{u}})\cdot\omega\,\,\,\,\,\text{.}%
\label{3.47}%
\end{equation}

If we express the frequencies as $\overline{\omega}=2\cdot\pi\cdot
\overline{\nu}$ and $\omega=2\cdot\pi\cdot\nu$ we obtain
\begin{align}
\overline{\nu}  & >\nu:\overline{\nu}=(1+\frac{l_{v_{z}}}{l_{u}})\cdot
\nu\,\,\,\,\,\,\text{,}\label{3.48}\\
\overline{\nu}  & <\nu:\overline{\nu}=(1-\frac{l_{v_{z}}}{l_{u}})\cdot
\nu\,\,\,\,\,\,\,\text{.}\label{3.49}%
\end{align}

The last relations represents a generalization of the \textit{standard
(longitudinal) Doppler effect} in $(\overline{L}_{n},g)$-spaces.

(b) If $l_{v_{z}}=0$ and the emitter moves only around an observer (detector)
with the Coriolis velocity $v_{c}=\pm l_{v_{c}}\cdot m_{\perp}$, i.e. if
$\alpha=0$, $\beta=\pm1$, then
\begin{equation}
\overline{\omega}=(1\pm\frac{l_{v_{c}}}{l_{u}})\cdot\omega\,\,\,\,\text{.}%
\label{3.50}%
\end{equation}

For $\overline{\omega}>\omega$%
\begin{equation}
\overline{\omega}=(1+\frac{l_{v_{c}}}{l_{u}})\cdot\omega\,\,\,\,\,\,\text{.}%
\label{3.51}%
\end{equation}

For $\overline{\omega}<\omega$%
\begin{equation}
\overline{\omega}=(1-\frac{l_{v_{c}}}{l_{u}})\cdot\omega\,\,\,\,\,\text{.}%
\label{3.52}%
\end{equation}

If we express the frequencies as $\overline{\omega}=2\cdot\pi\cdot
\overline{\nu}$ and $\omega=2\cdot\pi\cdot\nu$ we obtain
\begin{align}
\overline{\nu}  & >\nu:\overline{\nu}=(1+\frac{l_{v_{c}}}{l_{u}})\cdot
\nu\,\,\,\,\,\,\text{,}\label{3.53}\\
\overline{\nu}  & <\nu:\overline{\nu}=(1-\frac{l_{vc}}{l_{u}})\cdot
\nu\,\,\,\,\,\,\,\text{.}\label{3.54}%
\end{align}

The last relations represent a generalization of the \textit{transversal
Doppler effect} in $(\overline{L}_{n},g)$-spaces. The relations have the same
forms as these for the standard (longitudinal) Doppler effect but the
direction of the emitted signals changes in the time in contrast to the
standard Doppler effect. In the expressions for the standard Dopller effect
only the centrifugal (centripetal) velocity $v_{z}$ is replaced with the
Coriolis velocity $v_{c}$ for receiving the relations for the transversal
Doppler effect.

(c) If $l_{v_{z}}\neq0$ and $l_{v_{c}}\neq0$ then we have an accumulation of
both types of the Doppler effect
\[
\overline{\omega}=[1-(\alpha\cdot\frac{l_{v_{z}}}{l_{u}}+\beta\cdot
\frac{l_{v_{c}}}{l_{u}})]\cdot\omega\,\,\,\,\,\text{.}
\]

We have considered the Doppler effects without taking into account the
structures of the centrifugal (centripetal) velocity and of the Coriolis velocity.

It should be stressed that the generalized Doppler effects are a result of
pure kinematic considerations of the properties of a null (isotropic) vector
field by means of the kinematic characteristics of the relative velocity in
spaces with affine connections and metrics.

\subsection{Hubble's effect}

The Hubble's law has been considered on the grounds of the structures of the
centrifugal (centripetal) velocity \cite{Manoff-6}. Its connection to the
change of the frequency of an emitter could be found if we take into account
the structures of the centrifugal (centripetal) velocity and the Coriolis velocity.

Let us now consider the standard (longitudinal) Doppler effect when the
emitter has only centrifugal (centripetal) velocity with respect to the
observer. Then $\alpha=\pm1$, $\beta=0$ and
\[
\overline{\omega}=(1\pm\frac{l_{v_{z}}}{l_{u}})\cdot\omega\,\,\,\,\text{.}
\]

Since $\pm l_{v_{z}}=H\cdot l_{\xi_{\perp}}$, $v_{z}=H\cdot l_{\xi_{\perp}%
}\cdot n_{\perp}$ we obtain for the frequency $\overline{\omega}$ of the
emitter
\begin{equation}
\overline{\omega}=\omega+H\cdot\frac{l_{\xi_{\perp}}}{l_{u}}\cdot
\omega\,\,\,\,\,\,\text{.}\label{3.55}%
\end{equation}

The measured by the observer signals with frequency $\omega$ and the emitted
by the emitter signals with frequency $\overline{\omega}$ are related to each
other by the Hubble function $H$ and on this basis to the Hubble law.
Therefore, the radiation with frequency $\overline{w}$ by the emitter could be
expressed as
\begin{equation}
\overline{\omega}=(1+H\cdot\frac{l_{\xi_{\perp}}}{l_{u}})\cdot\omega
\label{3.56}%
\end{equation}
and it will be detected by the observer as radiation with frequency $\omega$.
The relative difference between both the frequencies (emitted $\overline
{\omega}$ and detected $\omega$)
\[
\frac{\overline{\omega}-\omega}\omega
\]
appears in the form
\begin{equation}
\frac{\overline{\omega}-\omega}\omega=H\cdot\frac{l_{\xi_{\perp}}}{l_{u}%
}\,\,\,\,\,\,\text{.}\label{3.57}%
\end{equation}

If we introduce the abbreviation
\begin{equation}
z=\frac{\overline{\omega}-\omega}\omega\label{3.58}%
\end{equation}
we obtain the relation between the emitted frequency $\overline{\omega}$ and
the frequency $\omega$ detected by the observer in the form
\begin{equation}
\frac{\overline{\omega}-\omega}\omega=z\,\,\,\,\,\,\,\,\text{,\thinspace
\thinspace\thinspace\thinspace\thinspace\thinspace\thinspace\thinspace
\thinspace\thinspace\thinspace}\overline{\omega}=(1+z)\cdot\omega
\,\,\,\,\,\,\,\,\text{,\thinspace\thinspace\thinspace\thinspace\thinspace
\thinspace\thinspace\thinspace\thinspace\thinspace\thinspace\thinspace
\thinspace\thinspace\thinspace\thinspace}z=H\cdot\frac{l_{\xi_{\perp}}}{l_{u}%
}\,\,\,\,\,\,\text{.}\label{3.59}%
\end{equation}

The change of the frequency $\overline{\omega}$ under the motion of the
emitter with centrifugal (centripetal) velocity $v_{z}$ relative to an
observer is called \textit{Hubble's effect}.

The quantity $z$ could be denoted as \textit{observed Hubble's shift frequency
parameter}. If $z=0$ then $H=0$, $v_{z}=0$, and there will be no difference
between the emitted and the detected frequencies, i.e. $\overline{\omega
}=\omega$, i.e. for $z=0$, it follows that $\overline{\omega}=\omega$. If we
take into account the explicit form of the Hubble function
\begin{equation}
H=\frac1{n-1}\cdot\theta\mp\sigma(n_{\perp},n_{\perp})\text{ ,}\label{3.60}%
\end{equation}
then $z=0$ leading to $H=0$ will be the case if
\begin{equation}
\theta=\pm(n-1)\cdot\sigma(n_{\perp},n_{\perp})\,\,\,\,\,\,\,\,\,\text{.}%
\label{3.61}%
\end{equation}

If $z>0$ the observed Hubble shift frequency parameter is called
\textit{Hubble's red shift}. If $z<0$ the observed Hubble's shift frequency
parameter is called \textit{Hubble's blue shift}. If $\overline{\omega}$ and
$\omega$ are known the observed Hubble'ss shift frequency parameter $z$ could
be found. If $\omega$ and $z$ are given then the corresponding $\overline
{\omega}$ could be estimated.

On the other side, from the explicit form of $z$%
\begin{equation}
z=H\cdot\frac{l_{\xi_{\perp}}}{l_{u}}=[\frac1{n-1}\cdot\theta\mp
\sigma(n_{\perp},n_{\perp})]\cdot\frac{l_{\xi_{\perp}}}{l_{u}}\label{5.24}%
\end{equation}
we could find the relation between the observed shift frequency parameter $z$
and the kinematic characteristics of the relative velocity such as the
expansion and shear velocities.

\textit{Special case}: $(\overline{L}_{n},g)$-spaces with shear-free relative
velocity: $\sigma:=0$.
\begin{equation}
z=\frac1{n-1}\cdot\theta\cdot\frac{l_{\xi_{\perp}}}{l_{u}}\text{
\thinspace\thinspace\thinspace\thinspace\thinspace\thinspace,\thinspace
\thinspace\thinspace\thinspace\thinspace\thinspace\thinspace\thinspace
\thinspace\thinspace\thinspace\thinspace\thinspace\thinspace}H=\frac
1{n-1}\cdot\theta\,\,\,\,\,\,\,\,\text{.}\label{5.25}%
\end{equation}

\textit{Special case}: $(\overline{L}_{n},g)$-spaces with expansion-free
relative velocity: $\theta:=0$.
\begin{equation}
z=\mp\sigma(n_{\perp},n_{\perp})\cdot\frac{l_{\xi_{\perp}}}{l_{u}%
}\,\,\,\,\,\,\,\,\,\text{,\thinspace\thinspace\thinspace\thinspace
\thinspace\thinspace\thinspace\thinspace\thinspace\thinspace\thinspace}%
H=\mp\sigma(n_{\perp},n_{\perp})\,\,\,\,\,\,\,\text{.}\label{5.26}%
\end{equation}

On the grounds of the observed shift frequency parameter $z$ the distance (the
length $l_{\xi_{\perp}}$ of $\xi_{\perp}$) between the observer [with the
world line $x^{i}(\tau)$ and velocity vector field $u=\frac d{d\tau}$] and the
observed object (at a distance $l_{\xi_{\perp}}$ from the observer) emitting
radiation with null vector field $k$ could be found as
\begin{equation}
l_{\xi_{\perp}}=z\cdot\frac{l_{u}}H=\frac{\overline{\omega}-\omega}\omega
\cdot\frac{l_{u}}H\,\,\,\,\,\,\text{.}\label{5.27}%
\end{equation}

On the other side, if $z$, $H$, and $l_{\xi_{\perp}}$ are known the absolute
value $l_{u}$ of the velocity vector $u$ could be found as
\begin{equation}
l_{u}=\frac Hz\cdot l_{\xi_{\perp}}=\frac{H\cdot\omega}{\overline{\omega
}-\omega}\cdot l_{\xi_{\perp}}\text{\thinspace\thinspace\thinspace
\thinspace\thinspace\thinspace\thinspace\thinspace.}\label{5.28}%
\end{equation}

\textit{Remark}. In the Einstein theory of gravitation (ETG) the absolute
value of $u$ is usually normalized to $1$ or $c$, i.e. $l_{u}=1,c$. Then the
last expression could be used for experimental check up of the velocity $c$ of
light in vacuum if $z$, $H$, and $l_{\xi_{\perp}}$ are known
\begin{equation}
c=\frac Hz\cdot l_{\xi_{\perp}}=\frac{H\cdot\omega}{\overline{\omega}-\omega
}\cdot l_{\xi_{\perp}}\,\,\,\,\,\,\text{.}\label{5.29}%
\end{equation}

Since
\[
z=[\frac1{n-1}\cdot\theta\mp\sigma(n_{\perp},n_{\perp})]\cdot\frac
{l_{\xi_{\perp}}}{l_{u}}=\frac{\overline{\omega}-\omega}\omega
\]
it follows that
\begin{equation}
l_{u}=\frac\omega{\overline{\omega}-\omega}\cdot[\frac1{n-1}\cdot\theta
\mp\sigma(n_{\perp},n_{\perp})]\cdot l_{\xi_{\perp}}\,\,\,\,\,\,\text{.}%
\label{5.30}%
\end{equation}

Analogous expression we can find for the length $l_{\xi_{\perp}}$ of the
vector field $\xi_{\perp}$%
\begin{equation}
l_{\xi_{\perp}}=(n-1)\cdot(\frac{\overline{\omega}}\omega-1)\cdot\frac{l_{u}%
}{\theta\mp(n-1)\cdot\sigma(n_{\perp},n_{\perp})}\,\,\,\,\,\,\,\text{.}%
\label{5.31}%
\end{equation}

\textit{Special case}: $(\overline{L}_{n},g)$-space with shear-free relative
velocity: $\sigma:=0$.
\begin{equation}
l_{u}=\frac\omega{\overline{\omega}-\omega}\cdot\frac1{n-1}\cdot\theta\cdot
l_{\xi_{\perp}}\,\,\,\,\,\,\text{,}\label{5.32}%
\end{equation}
\begin{equation}
l_{\xi_{\perp}}=(n-1)\cdot(\frac{\overline{\omega}}\omega-1)\cdot\frac{l_{u}%
}\theta\,\,\,\,\,\,\,\,\text{.}\label{5.33}%
\end{equation}

\textit{Special case}: $(\overline{L}_{n},g)$-space with expansion-free
relative velocity: $\theta:=0$.
\begin{equation}
l_{u}=\mp\frac\omega{\overline{\omega}-\omega}\cdot\sigma(n_{\perp},n_{\perp
})\cdot l_{\xi_{\perp}}\,\,\,\,\,\,\,\,\text{,}\label{5.34}%
\end{equation}
\begin{equation}
l_{\xi_{\perp}}=\mp(\frac{\overline{\omega}}\omega-1)\cdot\frac{l_{u}}%
{\sigma(n_{\perp},n_{\perp})}\,\,\,\,\,\,\,\,\,\,\text{.}\label{5.35}%
\end{equation}

By the use of the relation between the Hubble function $H$ and the observed
shift parameter $z$ we can express the centrifugal (centripetal) velocity by
means of the frequencies $\overline{\omega}$ and $\omega$. From
\[
v_{z}=H\cdot l_{\xi_{\perp}}\cdot n_{\perp}\text{\thinspace\thinspace
\thinspace\thinspace\thinspace\thinspace\thinspace,\thinspace\thinspace
\thinspace\thinspace\thinspace\thinspace\thinspace\thinspace\thinspace
\thinspace\thinspace\thinspace\thinspace\thinspace\thinspace\thinspace
}H=z\cdot\frac{l_{u}}{l_{\xi_{\perp}}}=(\frac{\overline{\omega}}\omega
-1)\cdot\frac{l_{u}}{l_{\xi_{\perp}}}\text{ \thinspace\thinspace\thinspace,}
\]
it follows that
\begin{equation}
v_{z}=z\cdot l_{u}\cdot n_{\perp}=(\frac{\overline{\omega}}\omega-1)\cdot
l_{u}\cdot n_{\perp}\,\,\,\,\,\,\text{.}\label{5.36}%
\end{equation}

Then
\[
g(v_{z},v_{z})=\mp(\frac{\overline{\omega}}\omega-1)^{2}\cdot l_{u}^{2}=\mp
l_{v_{z}}^{2}\text{ \thinspace\thinspace\thinspace\thinspace\thinspace,}
\]
\begin{align}
l_{v_{z}}  & =\pm(\frac{\overline{\omega}}\omega-1)\cdot l_{u}\text{
\thinspace\thinspace\thinspace\thinspace\thinspace\thinspace\thinspace
,\thinspace\thinspace\thinspace\thinspace\thinspace\thinspace\thinspace
\thinspace\thinspace\thinspace\thinspace}\pm l_{v_{z}}=(\frac{\overline
{\omega}}\omega-1)\cdot l_{u}\,\,\,\,\,\text{,}\label{5.37}\\
l_{v_{z}}  & >0\text{ \thinspace\thinspace\thinspace,\thinspace\thinspace
\thinspace\thinspace\thinspace\thinspace\thinspace\thinspace\thinspace}%
l_{u}>0\text{,\thinspace\thinspace\thinspace\thinspace\thinspace
\thinspace\thinspace\thinspace\thinspace}l_{\xi_{\perp}}%
>0\,\,\,\,\,\,\,\text{,}\label{3.37a}%
\end{align}
where (since $l_{v_{z}}>0$)
\begin{align}
\overline{\omega}  & >\omega:l_{v_{z}}=(\frac{\overline{\omega}}\omega-1)\cdot
l_{u}\text{\thinspace\thinspace\thinspace\thinspace\thinspace\thinspace
\thinspace\thinspace\thinspace\thinspace,}\label{5.38}\\
\overline{\omega}  & <\omega:l_{v_{z}}=(1-\frac{\overline{\omega}}\omega)\cdot
l_{u}\,\,\,\,\,\,\,\,\,\text{.}\label{5.39}%
\end{align}

\subsubsection{Einstein's theory of gravitation and the Hubble effect}

In Einstein's theory of gravitation it is assumed that $l_{u}=c$ and for all
other velocities $l_{v_{z}}\leq c$,\thinspace\thinspace$l_{v_{c}}\leq c$. From
the above expressions, it follows that for
\begin{align*}
\overline{\omega}  & >\omega:l_{v_{z}}=(\frac{\overline{\omega}}\omega-1)\cdot
l_{u}\leq l_{u}\,\,\,\text{,}\\
\frac{\overline{\omega}}\omega-1  & \leq1\,\,\,\,\text{,\thinspace
\thinspace\thinspace\thinspace\thinspace\thinspace\thinspace}l_{u}%
>0\,\,\,\,\,\text{,}%
\end{align*}
i.e. the emitted frequency $\overline{\omega}$ should be smaller than
$2\cdot\omega$%
\[
\frac{\overline{\omega}}\omega\leq2\,\,\,\,\,\,\,\text{,\thinspace
\thinspace\thinspace\thinspace\thinspace\thinspace\thinspace\thinspace
}\overline{\omega}\leq2\cdot\omega\,\,\,\,\,\,\text{.}
\]

Then for $\overline{\omega}>\omega$ the observed Hubble shift frequency
parameter $z\leq1$. For
\begin{align*}
\overline{\omega}  & <\omega:l_{v_{z}}=(1-\frac{\overline{\omega}}\omega)\cdot
l_{u}\leq l_{u}\,\,\,\,\,\text{,}\\
1-\frac{\overline{\omega}}\omega & \leq1:-\frac{\overline{\omega}}\omega
\leq0:\frac{\overline{\omega}}\omega\geq0\,\,\,\,\,\text{,}%
\end{align*}
i.e. $l_{v_{z}}\leq l_{u}$ is always fulfilled because of $\overline{\omega
}\geq0$ and $\omega\geq0$. Therefore, if the Einstein theory of gravitation is
a correct theory for description of the gravitational interaction under the
assumption that the absolute value $l_{u}$ of the velocity of the light
propagation is equal the absolute value $c$ of the velocity of light in
vacuum, then the red shift $z$ could not be bigger that $1$, i.e. $z\leq1$ for
$\overline{\omega}>\omega$. Such a limit for $z$ does not exist if
$\overline{\omega}<\omega$. If we could find experimentally that in some cases
$z>1$ then we should look for the reasons for this diversion from the theory
of the Hubble effect based on the kinematic characteristics of the relative
velocity. One of the possible reasons could be the influence of the
aberration's effect on the Hubble effect. The aberration effect is considered below.

\subsection{Aberration's effect}

The aberration effect is related the transversal Doppler effect in analogous
way as the Hubble effect is related to the standard (longitudinal) Doppler effect.

Let us now consider the transversal Doppler effect when the emitter has only
Coriolis' velocity with respect to the observer. Then $\alpha=0$, $\beta=\pm
1$, and
\begin{equation}
\overline{\omega}=\omega\pm\frac{l_{v_{c}}}{l_{u}}\cdot\omega
\,\,\,\,\,\,\,\,\,\,\,\text{,\thinspace\thinspace\thinspace\thinspace
\thinspace\thinspace\thinspace\thinspace\thinspace\thinspace\thinspace
\thinspace\thinspace\thinspace\thinspace\thinspace}v_{c}=\pm l_{v_{c}}\cdot
m_{\perp}\,\,\,\,\,\text{.}\label{3.62}%
\end{equation}

The Coriolis velocity $v_{c}=\pm l_{v_{c}}\cdot m_{\perp}$ could be
represented in the form \cite{Manoff-6}
\begin{align}
v_{c}  & =l_{\xi_{\perp}}\cdot\overline{g}[\sigma(n_{\perp})]\pm
\sigma(n_{\perp},n_{\perp})\cdot l_{\xi_{\perp}}\cdot n_{\perp}+l_{\xi_{\perp
}}\cdot\overline{g}[\omega(n_{\perp})]=\label{3.63}\\
& =l_{\xi_{\perp}}\cdot\overline{v}_{c}\text{ \thinspace,}\label{3.64}%
\end{align}
where
\begin{equation}
\overline{v}_{c}=\overline{g}[\sigma(n_{\perp})]\pm\sigma(n_{\perp},n_{\perp
})\cdot n_{\perp}+\overline{g}[\omega(n_{\perp})]\,\,\,\,\,\text{.}%
\label{3.65}%
\end{equation}

Then
\begin{align*}
v_{c}  & =l_{\xi_{\perp}}\cdot\overline{v}_{c}=\pm l_{v_{c}}\cdot m_{\perp
}\,\,\,\,\,\,\text{,}\\
\overline{v}_{c}  & =\pm l_{\overline{v}_{c}}\cdot m_{\perp}\,\,\,\,\text{,}\\
g(\overline{v}_{c},\overline{v}_{c})  & =\mp l_{\overline{v}_{c}}%
^{2}\,\,\,\,\,\,\text{,}%
\end{align*}
\begin{align}
v_{c}  & =\pm l_{v_{c}}\cdot m_{\perp}=l_{\xi_{\perp}}\cdot\overline{v}%
_{c}=l_{\xi_{\perp}}\cdot(\pm l_{\overline{v}_{c}}\cdot m_{\perp})=\pm
l_{\overline{v}_{c}}\cdot l_{\xi_{\perp}}\cdot m_{\perp}=\label{3.66}\\
& =H_{c}\cdot l_{\xi_{\perp}}\cdot m_{\perp}\,\,\,\text{,}\label{3.67}\\
v_{c}  & =\pm l_{\xi_{\perp}}\cdot l_{\overline{v}_{c}}\cdot m_{\perp}=\pm
l_{v_{c}}\cdot m_{\perp}\text{ \thinspace\thinspace,\thinspace\thinspace
\thinspace\thinspace\thinspace\thinspace\thinspace\thinspace\thinspace
\thinspace}\pm l_{v_{c}}=\pm l_{\xi_{\perp}}\cdot l_{\overline{v}_{c}%
}\,\,\,\,\text{,}\label{3.67a}\\
l_{v_{c}}  & =l_{\xi_{\perp}}\cdot l_{\overline{v}_{c}}%
\,\,\,\,\,\text{,\thinspace\thinspace\thinspace\thinspace\thinspace
\thinspace\thinspace\thinspace}g(v_{c},v_{c})=v_{c}^{2}=\mp l_{v_{c}}^{2}=\mp
l_{\xi_{\perp}}^{2}\cdot l_{\overline{v}_{c}}^{2}\,\,\,\,\,\,\,\text{.}%
\label{3.67b}%
\end{align}
where
\begin{equation}
H_{c}=\pm l_{\overline{v}_{c}}\text{ \thinspace\thinspace\thinspace
\thinspace,\thinspace\thinspace\thinspace\thinspace\thinspace\thinspace
\thinspace\thinspace\thinspace\thinspace\thinspace\thinspace}\pm l_{v_{c}%
}=H_{c}\cdot l_{\xi_{\perp}}\,\,\,\,\,\text{.}\label{3.68}%
\end{equation}

Then $v_{c}=\pm l_{\overline{v}_{c}}\cdot l_{\xi_{\perp}}\cdot m_{\perp}%
=H_{c}\cdot l_{\xi_{\perp}}\cdot m_{\perp}$ (compare with $v_{z}=H\cdot
l_{\xi_{\perp}}\cdot n_{\perp}$).

After introducing the last (previous) expression for $l_{v_{c}}$ into the
relation for $\overline{\omega}$, it follows that
\begin{equation}
\overline{\omega}=\omega\pm\frac{l_{v_{c}}}{l_{u}}\cdot\omega=\omega
+H_{c}\cdot\frac{l_{\xi_{\perp}}}{l_{u}}\cdot\omega
\,\,\,\,\,\,\,\,\,\text{,\thinspace\thinspace\thinspace\thinspace
\thinspace\thinspace\thinspace\thinspace\thinspace\thinspace\thinspace
\thinspace\thinspace}H_{c}=\pm l_{\overline{v}_{c}}\,\,\,\,\,\text{.}%
\label{3.69}%
\end{equation}

The (invariant) function $H_{c}$ is called \textit{Coriolis' function}.

For the emitted frequency $\overline{\omega}$ we obtain an analogous
expression as in the case of the Hubble effect (only $H$ is replaced by
$H_{c}$)
\begin{equation}
\overline{\omega}=(1+H_{c}\cdot\frac{l_{\xi_{\perp}}}{l_{u}})\cdot
\omega\,\,\,\,\text{.}\label{3.70}%
\end{equation}

The detected by the observer frequency is $\omega$ if the emitter is moving
with a Coriolis velocity $v_{c}$ around the observer and emitting signals with
a frequency $\overline{\omega}$. The relative difference between both the
frequencies (emitted $\overline{\omega}$ and detected $\omega$)
\[
\frac{\overline{\omega}-\omega}\omega
\]
appears in the form
\begin{equation}
\frac{\overline{\omega}-\omega}\omega=H_{c}\cdot\frac{l_{\xi_{\perp}}}{l_{u}%
}\,\,\,\,\,\,\text{.}\label{3.71}%
\end{equation}

If we introduce the abbreviation
\begin{equation}
z_{c}=\frac{\overline{\omega}-\omega}\omega\label{3.72}%
\end{equation}
we obtain the relation between the emitted frequency $\overline{\omega}$ and
the frequency $\omega$ detected by the observer in the form
\begin{equation}
\frac{\overline{\omega}-\omega}\omega=z_{c}\,\,\,\,\,\,\,\,\text{,\thinspace
\thinspace\thinspace\thinspace\thinspace\thinspace\thinspace\thinspace
\thinspace\thinspace\thinspace}\overline{\omega}=(1+z_{c})\cdot\omega
\,\,\,\,\,\,\,\,\text{,\thinspace\thinspace\thinspace\thinspace\thinspace
\thinspace\thinspace\thinspace\thinspace\thinspace\thinspace\thinspace
\thinspace\thinspace\thinspace\thinspace}z_{c}=H_{c}\cdot\frac{l_{\xi_{\perp}%
}}{l_{u}}=\pm\frac{l_{v_{c}}}{l_{u}}\,\,\,\,\,\,\text{.}\label{3.73}%
\end{equation}

The change of the frequency $\overline{\omega}$ under the motion of the
emitter with Coriolis' velocity $v_{c}$ relative to an observer is called
\textit{aberration's effect}.

The quantity $z_{c}$ could be denoted as \textit{observed aberration's shift
frequency parameter}. If $z_{c}=0$ then $H_{c}=0$, $v_{c}=0$, and there will
be no difference between the emitted and the detected frequencies, i.e.
$\overline{\omega}=\omega$, i.e. for $z_{c}=0$, it follows that $\overline
{\omega}=\omega$.

If $z_{c}>0$ the observed aberration shift frequency parameter is called
\textit{aberration's red shift}. If $z_{c}<0$ the observed aberration shift
frequency parameter is called \textit{aberration's blue shift}. If
$\overline{\omega}$ and $\omega$ are known the observed aberration shift
frequency parameter $z_{c}$ could be found. If $\omega$ and $z_{c}$ are given
then the corresponding $\overline{\omega}$ could be estimated.

On the other side, from the explicit form of $z_{c}$%
\begin{equation}
z_{c}=H_{c}\cdot\frac{l_{\xi_{\perp}}}{l_{u}}\,\,\,\,\,\,\,\,\text{,\thinspace
\thinspace\thinspace\thinspace\thinspace\thinspace\thinspace\thinspace
\thinspace\thinspace\thinspace\thinspace\thinspace\thinspace\thinspace}%
H_{c}=\pm l_{\overline{v}_{c}}\,\,\,\,\,\text{,}\label{3.74}%
\end{equation}
we could find the relation between the observed shift frequency parameter
$z_{c}$ and the kinematic characteristics of the relative velocity such as the
shear and rotation velocities.

Since
\begin{align}
v_{c}^{2}  & =\mp l_{v_{c}}^{2}=\mp l_{\xi_{\perp}}^{2}\cdot l_{\overline
{v}_{c}}^{2}=\mp l_{\xi_{\perp}}\cdot H_{c}^{2}=\label{3.75}\\
& =l_{\xi_{\perp}}\cdot\{\overline{g}(\sigma(n_{\perp}),\sigma(n_{\perp}%
))\pm[\sigma(n_{\perp},n_{\perp})]^{2}+\overline{g}(\omega(n_{\perp}%
),\omega(n_{\perp}))+\nonumber\\
& +2\cdot\overline{g}(\sigma(n_{\perp}),\omega(n_{\perp}))\}\,\,\,\,\text{,}%
\label{3.76}%
\end{align}
it follows that
\begin{align}
H_{c}^{2}  & =\mp\{\overline{g}(\sigma(n_{\perp}),\sigma(n_{\perp}))\pm
[\sigma(n_{\perp},n_{\perp})]^{2}+\overline{g}(\omega(n_{\perp}),\omega
(n_{\perp}))+\nonumber\\
& +2\cdot\overline{g}(\sigma(n_{\perp}),\omega(n_{\perp}))\}\,\,\,\,\text{,}%
\label{3.77}%
\end{align}
and
\begin{align}
H_{c}  & =\pm\{\mp\{\overline{g}(\sigma(n_{\perp}),\sigma(n_{\perp}%
))\pm[\sigma(n_{\perp},n_{\perp})]^{2}+\overline{g}(\omega(n_{\perp}%
),\omega(n_{\perp}))+\nonumber\\
& +2\cdot\overline{g}(\sigma(n_{\perp}),\omega(n_{\perp}))\}\}^{1/2}%
\text{\thinspace\thinspace\thinspace\thinspace.}\label{3.78}%
\end{align}

\textit{Remark}. The quantity $\omega$ in the structure of $H_{c}$ and $v_{c}$
is an antisymmetric covariant tensor of second rank $\omega\in\otimes
_{2}(M)=\Lambda^{2}(M)$ interpreted as the rotation velocity tensor. It should
be distinguished from the frequency $\omega$ detected by the
observer.\thinspace The tensor $\omega$ appears only in the structure of
$H_{c}$ and $z_{c}$ in contrast to the frequency $\omega$ appearing out of
these quantities.

\textit{Special case}: $(\overline{L}_{n},g)$-spaces with shear-free relative
velocity: $\sigma:=0$.
\begin{equation}
z_{c}=\pm[\mp\overline{g}(\omega(n_{\perp}),\omega(n_{\perp}))]^{1/2}%
\cdot\frac{l_{\xi_{\perp}}}{l_{u}}\text{ \thinspace\thinspace\thinspace
\thinspace\thinspace\thinspace,\thinspace\thinspace\thinspace\thinspace
\thinspace\thinspace\thinspace\thinspace\thinspace\thinspace\thinspace
\thinspace\thinspace\thinspace}H_{c}=\pm[\mp\overline{g}(\omega(n_{\perp
}),\omega(n_{\perp}))]^{1/2}\,\,\,\,\,\,\,\,\text{.}\label{3.79}%
\end{equation}

\textit{Special case}: $(\overline{L}_{n},g)$-spaces with rotation-free
relative velocity: $\omega:=0$.
\begin{align}
z_{c}  & =\pm\{\mp\overline{g}(\sigma(n_{\perp}),\sigma(n_{\perp}%
))-[\sigma(n_{\perp},n_{\perp})]^{2}\}^{1/2}\cdot\frac{l_{\xi_{\perp}}}{l_{u}%
}\,\,\,\,\,\,\,\,\,\text{,\thinspace\thinspace\thinspace\thinspace
\thinspace\thinspace\thinspace\thinspace\thinspace\thinspace\thinspace
}\label{3.80}\\
H_{c}  & =\pm\{\mp\overline{g}(\sigma(n_{\perp}),\sigma(n_{\perp}%
))-[\sigma(n_{\perp},n_{\perp})]^{2}\}^{1/2}\,\,\,\,\,\,\,\text{.}\label{3.81}%
\end{align}

On the grounds of the observed aberration shift frequency parameter $z_{c}$
the distance (the length $l_{\xi_{\perp}}$ of $\xi_{\perp}$) between the
observer [with the world line $x^{i}(\tau)$ and velocity vector field $u=\frac
d{d\tau}$] and the observed object (at a distance $l_{\xi_{\perp}} $ from the
observer) emitting radiation with null vector field $\widetilde{k} $ could be
found as
\begin{equation}
l_{\xi_{\perp}}=z_{c}\cdot\frac{l_{u}}{H_{c}}=\frac{\overline{\omega}-\omega
}\omega\cdot\frac{l_{u}}{H_{c}}\,\,\,\,\,\,\text{.}\label{3.82}%
\end{equation}

On the other side, if $z_{c}$, $H_{c}$, and $l_{\xi_{\perp}}$ are known the
absolute value $l_{u}$ of the velocity vector $u$ could be found as
\begin{equation}
l_{u}=\frac{H_{c}}{z_{c}}\cdot l_{\xi_{\perp}}=\frac{H_{c}\cdot\omega
}{\overline{\omega}-\omega}\cdot l_{\xi_{\perp}}\text{\thinspace
\thinspace\thinspace\thinspace\thinspace\thinspace\thinspace\thinspace
.}\label{3.84}%
\end{equation}

\textit{Remark}. In the Einstein theory of gravitation (ETG) the absolute
value of $u$ is usually normalized to $1$ or $c$, i.e. $l_{u}=1,c$. The last
(previous) expression for $l_{u}$ could be used for experimental check up of
the velocity $c$ of light in vacuum if $z_{c}$, $H_{c}$, and $l_{\xi_{\perp}}$
are known
\begin{align}
c  & =\frac{H_{c}}{z_{c}}\cdot l_{\xi_{\perp}}=\frac{H_{c}\cdot\omega
}{\overline{\omega}-\omega}\cdot l_{\xi_{\perp}}=\label{3.85}\\
& =\pm\frac\omega{\overline{\omega}-\omega}\cdot\{\mp\{\overline{g}%
(\sigma(n_{\perp}),\sigma(n_{\perp}))\pm[\sigma(n_{\perp},n_{\perp}%
)]^{2}+\overline{g}(\omega(n_{\perp}),\omega(n_{\perp}))+\nonumber\\
& +2\cdot\overline{g}(\sigma(n_{\perp}),\omega(n_{\perp}))\}\}^{1/2}\cdot
l_{\xi_{\perp}}\,\,\,\text{.}\label{3.86}%
\end{align}

Since
\[
z_{c}=H_{c}\cdot\frac{l_{\xi_{\perp}}}{l_{u}}=\frac{\overline{\omega}-\omega
}\omega
\]
it follows that
\begin{equation}
l_{u}=\frac\omega{\overline{\omega}-\omega}\cdot H_{c}\cdot l_{\xi_{\perp}%
}=\frac{H_{c}}{z_{c}}\cdot l_{\xi_{\perp}}\text{.}\label{3.87}%
\end{equation}

Analogous expression could be found for the length $l_{\xi_{\perp}}$ of the
vector field $\xi_{\perp}$%
\begin{equation}
l_{\xi_{\perp}}=z_{c}\cdot\frac{l_{u}}{H_{c}}=(\frac{\overline{\omega}}%
\omega-1)\cdot\frac{l_{u}}{H_{c}}\,\,\,\,\,\text{.}\label{3.88}%
\end{equation}

\textit{Special case}: $(\overline{L}_{n},g)$-space with shear-free relative
velocity: $\sigma:=0$.
\begin{equation}
l_{u}=\pm\frac\omega{\overline{\omega}-\omega}\cdot[\mp\overline{g}%
(\omega(n_{\perp}),\omega(n_{\perp}))]^{1/2}\cdot l_{\xi_{\perp}%
}\,\,\,\,\,\,\text{,}\label{3.89}%
\end{equation}
\begin{equation}
l_{\xi_{\perp}}=\pm(\frac{\overline{\omega}}\omega-1)\cdot\,\,\,\frac{l_{u}%
}{[\mp\overline{g}(\omega(n_{\perp}),\omega(n_{\perp}))]^{1/2}}%
\,\,\,\,\,\text{.}\label{3.90}%
\end{equation}

\textit{Special case}: $(\overline{L}_{n},g)$-space with rotation-free
relative velocity: $\omega:=0$.
\begin{equation}
l_{u}=\frac{H_{c}}{z_{c}}\cdot l_{\xi_{\perp}}=\pm\frac\omega{\overline
{\omega}-\omega}\cdot\{\mp\overline{g}(\sigma(n_{\perp}),\sigma(n_{\perp
}))-[\sigma(n_{\perp},n_{\perp})]^{2}\}^{1/2}\cdot l_{\xi_{\perp}%
}\,\,\,\,\,\,\,\,\text{,}\label{3.91}%
\end{equation}
\begin{equation}
l_{\xi_{\perp}}=z_{c}\cdot\frac{l_{u}}{H_{c}}=\pm(\frac{\overline{\omega}%
}\omega-1)\cdot\frac{l_{u}}{\{\mp\overline{g}(\sigma(n_{\perp}),\sigma
(n_{\perp}))-[\sigma(n_{\perp},n_{\perp})]^{2}\}^{1/2}}%
\,\,\,\,\,\,\,\,\,\,\text{.}\label{3.92}%
\end{equation}

By the use of the relation between the Coriolis function $H_{c}$ and the
observed aberration shift parameter $z_{c}$ we can express the centrifugal
(centripetal) velocity $v_{c}$ by means of the frequencies $\overline{\omega}$
and $\omega$. From
\[
v_{c}=H_{c}\cdot l_{\xi_{\perp}}\cdot m_{\perp}\text{\thinspace\thinspace
\thinspace\thinspace\thinspace\thinspace\thinspace,\thinspace\thinspace
\thinspace\thinspace\thinspace\thinspace\thinspace\thinspace\thinspace
\thinspace\thinspace\thinspace\thinspace\thinspace\thinspace\thinspace}%
H_{c}=z_{c}\cdot\frac{l_{u}}{l_{\xi_{\perp}}}=(\frac{\overline{\omega}}%
\omega-1)\cdot\frac{l_{u}}{l_{\xi_{\perp}}}\text{ \thinspace\thinspace
\thinspace,}
\]
it follows that
\begin{equation}
v_{c}=z_{c}\cdot l_{u}\cdot m_{\perp}=(\frac{\overline{\omega}}\omega-1)\cdot
l_{u}\cdot m_{\perp}\,\,\,\,\,\,\text{.}\label{3.93}%
\end{equation}

Then
\[
g(v_{z},v_{z})=\mp(\frac{\overline{\omega}}\omega-1)^{2}\cdot l_{u}^{2}=\mp
l_{v_{z}}^{2}\text{ \thinspace\thinspace\thinspace\thinspace\thinspace,}
\]
\begin{align}
l_{v_{c}}  & =\pm(\frac{\overline{\omega}}\omega-1)\cdot l_{u}\text{
\thinspace\thinspace\thinspace\thinspace\thinspace\thinspace\thinspace
,\thinspace\thinspace\thinspace\thinspace\thinspace\thinspace\thinspace
\thinspace\thinspace\thinspace\thinspace}\pm l_{v_{c}}=(\frac{\overline
{\omega}}\omega-1)\cdot l_{u}\,\,\,\,\,\text{,}\label{3.94}\\
l_{v_{z}}  & >0\text{ \thinspace\thinspace\thinspace,\thinspace\thinspace
\thinspace\thinspace\thinspace\thinspace\thinspace\thinspace\thinspace}%
l_{u}>0\text{,\thinspace\thinspace\thinspace\thinspace\thinspace
\thinspace\thinspace\thinspace\thinspace}l_{\xi_{\perp}}%
>0\,\,\,\,\,\,\,\text{,}\label{3.95}%
\end{align}
where (since $l_{v_{c}}>0$)
\begin{align}
\overline{\omega}  & >\omega:l_{v_{c}}=(\frac{\overline{\omega}}\omega-1)\cdot
l_{u}\text{\thinspace\thinspace\thinspace\thinspace\thinspace\thinspace
\thinspace\thinspace\thinspace\thinspace,}\label{3.96}\\
\overline{\omega}  & <\omega:l_{v_{c}}=(1-\frac{\overline{\omega}}\omega)\cdot
l_{u}\,\,\,\,\,\,\,\,\,\text{.}\label{3.97}%
\end{align}

\textit{Remark}. Analogous considerations for the Einstein theory of
gravitation as in the case of the Hubble effect could be made also for the
case of the aberration effect. If $l_{v_{c}}\leq l_{u}$ then from
\[
z_{c}=\pm\frac{l_{\overline{v}_{c}}}{l_{u}}\,\,\,\,\text{,}
\]
it follows that

(a) for $z_{c}=+\frac{l_{\overline{v}_{c}}}{l_{u}}\leq\frac{l_{u}}{l_{u}}=1$,

(b) for $z_{c}=-\frac{l_{\overline{v}_{c}}}{l_{u}}\geq-1$: \thinspace
\thinspace\thinspace\thinspace$\frac{l_{\overline{v}_{c}}}{l_{u}}\leq1$.

If $l_{u}=c$ and $l_{v_{c}}\leq l_{u}$ then for the case $\overline{\omega
}>\omega$ we have the condition $z\leq1$. For $\overline{\omega}<\omega$ there
is no such condition because $l_{v_{c}}\leq l_{u}$ is fulfilled automatically.

\subsection{Accumulation of Hubble's effect and aberration's effect}

If the relative velocity $_{rel}v$ of the emitter is a superposition of the
centrifugal (centripetal) velocity $v_{z}$ and the Coriolis velocity $v_{c}$
with respect to the observer, i.e. if
\begin{equation}
_{rel}v=v_{z}+v_{c}\,\,\,\,\,\text{,\thinspace\thinspace\thinspace
\thinspace\thinspace\thinspace\thinspace\thinspace\thinspace}v_{z}%
\neq0\,\,\,\,\,\,\text{,\thinspace\thinspace\thinspace\thinspace
\thinspace\thinspace\thinspace}v_{c}\neq0\,\,\,\,\,\text{,}\label{3.98}%
\end{equation}
then both the Hubble and the aberration effects influence each other. From the
general expression for the emitted frequency $\overline{\omega}$ as function
of the detected frequency $\omega$%
\[
\overline{\omega}=[1-(\alpha\cdot\frac{l_{v_{z}}}{l_{u}}+\beta\cdot
\frac{l_{v_{c}}}{l_{u}})]\cdot\omega\,\,\,\,\text{,}
\]
after substituting $\pm l_{v_{z}}$ with $\pm l_{v_{z}}=H\cdot l_{\xi_{\perp}}%
$, $l_{v_{z}}=\pm H\cdot l_{\xi_{\perp}}$, and $\pm l_{v_{c}}=H_{c}\cdot
l_{\xi_{\perp}}$, \thinspace\thinspace$l_{v_{c}}=\pm H_{c}\cdot l_{\xi_{\perp
}}$, we obtain
\begin{equation}
\overline{\omega}=\omega\mp(\alpha\cdot H+\beta\cdot H_{c})\cdot\frac
{l_{\xi_{\perp}}}{l_{u}}\cdot\omega\,\,\,\,\,\text{,}\label{3.99}%
\end{equation}

\begin{equation}
\overline{\omega}=[1\mp(\alpha\cdot H+\beta\cdot H_{c})]\cdot\frac
{l_{\xi_{\perp}}}{l_{u}}\cdot\omega\,\,\,\,\,\,\,\,\text{.}\label{3.100}%
\end{equation}

The change of the frequency of the emitter is caused by both the velocities
$v_{z}$ and $v_{c}$. Instead of the Hubble function $H$ (Hubble's effect) or
of the aberration function $H_{c}$ (aberration's effect) a combination of both
functions appear in the expression for $\overline{\omega}$.

If we, further, express $H$ and $H_{c}$ with $z$ and $z_{c}$ respectively
\[
H=z\cdot\frac{l_{u}}{l_{\xi_{\perp}}}\,\,\,\,\,\,\,\text{,\thinspace
\thinspace\thinspace\thinspace\thinspace\thinspace\thinspace\thinspace
\thinspace}H_{c}=z_{c}\cdot\frac{l_{u}}{l_{\xi_{\perp}}}\,\,\,\,\,\text{,}
\]
we obtain for $\overline{\omega}$%
\begin{align}
\overline{\omega}  & =\omega\mp(\alpha\cdot z+\beta\cdot z_{c})\cdot
\frac{l_{u}}{l_{\xi_{\perp}}}\cdot\frac{l_{\xi_{\perp}}}{l_{u}}\cdot
\omega=\nonumber\\
& =\omega\,\mp(\alpha\cdot z+\beta\cdot z_{c})\cdot\omega\,\,\,\,\text{.}%
\label{3.101}%
\end{align}
and
\begin{equation}
\frac{\overline{\omega}-\omega}\omega=\mp(\alpha\cdot z+\beta\cdot
z_{c})\,\,\text{\thinspace\thinspace\thinspace\thinspace.}\label{3.102}%
\end{equation}

Both the Hubble effect and the aberration effect could compensate each other
if
\begin{equation}
\alpha\cdot z+\beta\cdot z_{c}=0\,\,\,\,\,\text{,}\label{3.103}%
\end{equation}
i.e. if
\begin{equation}
z_{c}=-\frac\alpha\beta\cdot z\,\,\,\,\,\text{,\thinspace\thinspace
}\label{3.104}%
\end{equation}
or if
\begin{align}
z_{c}  & =-\frac\alpha\beta\cdot z=H_{c}\cdot\frac{l_{\xi_{\perp}}}{l_{u}%
}=-\frac\alpha\beta\cdot H\cdot\frac{l_{\xi_{\perp}}}{l_{u}}%
\,\,\,\,\,\,\text{,}\label{3.105}\\
H_{c}  & =-\frac\alpha\beta\cdot H\,\,\,\,\,\text{.}\label{3.106}%
\end{align}

Under the above conditions [$z_{c}=-(\alpha/\beta)\cdot z$ or $H_{c}%
=-(\alpha/\beta)\cdot H$] there will be no change of the frequency
$\overline{\omega}$ of the emitter. The same frequency $\overline{\omega
}=\omega$ will also be detected by the observer.

Since $\alpha=\mp\,cos(\widetilde{n}_{\perp},n_{\perp})$ and $\beta=\mp
sin\,(\widetilde{n}_{\perp},n_{\perp})$ the relations for $z_{c}$ and $H_{c}$
will take the form
\begin{align}
z_{c}  & =-\frac\alpha\beta\cdot z=-\frac{cos\,(\widetilde{n}_{\perp}%
,n_{\perp})}{sin\,(\widetilde{n}_{\perp},n_{\perp})}\cdot H\cdot\frac
{l_{\xi_{\perp}}}{l_{u}}=\label{3.107}\\
& =-H\cdot\frac{l_{\xi_{\perp}}}{l_{u}}\cdot cotg(\widetilde{n}_{\perp
},n_{\perp})\,\,\,\,\,\,\,\,\text{,}\label{3.108}%
\end{align}
\begin{align}
H_{c}  & =-\frac\alpha\beta\cdot H=-\frac{cos\,(\widetilde{n}_{\perp}%
,n_{\perp})}{sin\,(\widetilde{n}_{\perp},n_{\perp})}\cdot H=\label{3.109}\\
& =-H\cdot cotg\,(\widetilde{n}_{\perp},n_{\perp})\,\,\,\,\text{.}%
\label{3.110}%
\end{align}

If $cotg\,(\widetilde{n}_{\perp},n_{\perp})=0$ then $z_{c}=0$, $H_{c}=0$, and
$\overline{\omega}=\omega$.

Denoting
\begin{equation}
\frac{\overline{\omega}-\omega}\omega=z_{gen}\,\,\,\,\,\text{,}\label{3.111}%
\end{equation}
we obtain
\begin{equation}
z_{gen}=\mp(\alpha\cdot z+\beta\cdot z_{c})\text{\thinspace\thinspace
\thinspace\thinspace\thinspace\thinspace\thinspace\thinspace\thinspace
\thinspace\thinspace.}\label{3.112}%
\end{equation}

The quantity $z_{gen}$ is the general observed shift parameter as a result of
both effects. Then
\begin{align}
z_{gen}  & =H_{gen}\cdot\frac{l_{\xi_{\perp}}}{l_{u}}=\mp(\alpha\cdot
z+\beta\cdot z_{c})\text{\thinspace}=\mp(\alpha\cdot H+\beta\cdot H_{c}%
)\cdot\frac{l_{\xi_{\perp}}}{l_{u}}\,=\label{3.113}\\
& =[H\cdot cos\,(\widetilde{n}_{\perp},n_{\perp})+H_{c}\cdot sin\,(\widetilde
{n}_{\perp},n_{\perp})]\cdot\frac{l_{\xi_{\perp}}}{l_{u}}=\label{3.115}\\
& =z\cdot cos\,(\widetilde{n}_{\perp},n_{\perp})+z_{c}\cdot sin\,(\widetilde
{n}_{\perp},n_{\perp})\,\,\,\,\,\,\,\text{.}\label{3.116}%
\end{align}
\begin{equation}
H_{gen}=\mp(\alpha\cdot H+\beta\cdot H_{c})\,\,\,\,\,\,\text{.}\label{3.117}%
\end{equation}

The general observed shift parameter $z_{gen}$ could take values
$z_{gen}\gtreqqless0$. These values depend on the motion of the emitter
relative to the observer. In some cases, when $z_{gen}=0$, the observer could
not find any difference between the emitted frequency $\overline{\omega}$ and
the detected frequency $\omega$ despite of the relative motion between emitter
and observer. In these cases, the effect of a motion with centrifugal
(centripetal) velocity of the emitter will be compensated by the effect,
generated by a motion with Coriolis' velocity.

From the relations
\[
z_{gen}=H_{gen}\cdot\frac{l_{\xi_{\perp}}}{l_{u}}%
\,\,\,\,\,\,\,\,\,\,\,\text{,\thinspace\thinspace\thinspace\thinspace
\thinspace\thinspace\thinspace\thinspace\thinspace\thinspace\thinspace
\thinspace\thinspace\thinspace\thinspace}z_{gen}=\frac{\overline{\omega
}-\omega}\omega\text{ \thinspace\thinspace\thinspace\thinspace,}
\]
we can find $l_{\xi_{\perp}}$ if $z_{gen}$, $H_{gen}$, and $l_{u}$ are given.
Then
\begin{equation}
l_{\xi_{\perp}}=\frac{z_{gen}}{H_{gen}}\cdot l_{u}=(\frac{\overline{\omega}%
}\omega-1)\cdot\frac1{H_{gen}}\cdot l_{u}\,\,\,\,\,\,\,\text{.}\label{3.118}%
\end{equation}

In the same way, if $z_{gen}$, $H_{gen}$, and $l_{\xi_{\perp}}$ are known the
absolute value $l_{u}$ of the velocity vector $u$ could be determined as
\begin{equation}
l_{u}=\frac{H_{gen}}{z_{gen}}\cdot l_{\xi_{\perp}}=\frac{H_{gen}\cdot\omega
}{\overline{\omega}-\omega}\cdot l_{\xi_{\perp}}\,\,\,\,\,\,\text{.}%
\label{3.119}%
\end{equation}
\begin{align}
z_{gen}  & =H_{gen}\cdot\frac{l_{\xi_{\perp}}}{l_{u}}=\mp(\alpha\cdot
H+\beta\cdot H_{c})\cdot\frac{l_{\xi_{\perp}}}{l_{u}}\,\,\,\,\,\,\,\,\text{,}%
\label{3.120}\\
l_{u}  & =\mp\frac{l_{\xi_{\perp}}}{z_{gen}}\cdot(\alpha\cdot H+\beta\cdot
H_{c})=\label{3.121}\\
& =-\frac1{z_{gen}}\cdot(\alpha\cdot l_{v_{z}}+\beta\cdot l_{v_{c}%
})\,\,\,\,\,\,\,\text{.}\label{3.122}%
\end{align}

\textit{Remark}. From the relation
\[
l_{u}=-\frac1{z_{gen}}\cdot(\alpha\cdot l_{v_{z}}+\beta\cdot l_{v_{c}%
})\,\,\,\,\,\,\,\,\,\,\,\text{,\thinspace\thinspace\thinspace\thinspace
\thinspace\thinspace\thinspace\thinspace\thinspace}l_{v_{z}}\geq
0\,\,\,\,\,\,\,\,\text{,\thinspace\thinspace\thinspace\thinspace
\thinspace\thinspace\thinspace\thinspace}l_{v_{c}}\geq0\,\,\,\,\,\,\text{,}
\]
under the condition that $l_{u}>0$, it follows that
\begin{equation}
-\frac1{z_{gen}}\cdot(\alpha\cdot l_{v_{z}}+\beta\cdot l_{v_{c}}%
)\,>0\,\,\,\,\,\,\text{.}\label{3.123}%
\end{equation}

Therefore, either $z_{gen}<0$ and $\alpha\cdot l_{v_{z}}+\beta\cdot l_{v_{c}%
}>0 $ or $z_{gen}>0$ and $\alpha\cdot l_{v_{z}}+\beta\cdot l_{v_{c}}<0$.

(a) For $z_{gen}>0$, it follows that
\begin{align}
\alpha\cdot l_{v_{z}}+\beta\cdot l_{v_{c}}  & <0\,\,\,\,\,\,\text{,\thinspace
\thinspace\thinspace\thinspace\thinspace\thinspace\thinspace\thinspace
\thinspace\thinspace\thinspace}\beta\cdot\text{\thinspace}l_{v_{c}}%
<-\alpha\cdot l_{v_{z}}\,\,\,\,\,\,\,\text{,\thinspace\thinspace
\thinspace\thinspace\thinspace\thinspace\thinspace\thinspace\thinspace
\thinspace}\alpha\cdot l_{v_{z}}<-\beta\cdot l_{v_{c}}\,\,\,\,\text{,}%
\nonumber\\
\alpha & >0:0<l_{v_{z}}<-\frac\beta\alpha\cdot l_{v_{c}}%
\,\,\,\text{,\thinspace\thinspace\thinspace\thinspace\thinspace\thinspace
\thinspace\thinspace\thinspace\thinspace\thinspace\thinspace}-\frac\beta
\alpha>0\,\,\,\,\,\,\text{,}\label{3.124}\\
\alpha & <0:l_{v_{z}}>-\frac\beta\alpha\cdot l_{v_{c}}\,\,\,\text{,\thinspace
\thinspace\thinspace\thinspace\thinspace\thinspace\thinspace}l_{v_{z}%
}>0\,\,\,\,\,\,\text{.}\label{3.125}%
\end{align}

(b) For $z_{gen}<0$, it follows that
\begin{align}
\alpha\cdot l_{v_{z}}+\beta\cdot l_{v_{c}}  &
>0\,\,\,\,\,\,\,\,\text{,\thinspace\thinspace\thinspace\thinspace
\thinspace\thinspace\thinspace\thinspace\thinspace}\,\alpha\cdot l_{v_{z}%
}>-\beta\cdot l_{v_{c}}\,\,\,\,\,\,\,\text{,}\nonumber\\
\alpha & >0:l_{v_{z}}>-\frac\beta\alpha\cdot l_{v_{c}}%
\,\,\,\,\,\,\,\text{,\thinspace\thinspace\thinspace\thinspace\thinspace
\thinspace\thinspace\thinspace}l_{v_{z}}>0\,\,\,\,\,\text{,}\label{3.126}\\
\alpha & <0:0<l_{v_{z}}<-\frac\beta\alpha\cdot l_{v_{c}}%
\,\,\,\,\,\,\,\text{,\thinspace\thinspace\thinspace\thinspace\thinspace
\thinspace\thinspace}-\frac\beta\alpha>0\,\,\,\,\text{.}\label{3.127}%
\end{align}

If, in addition, it is assumed that $l_{v_{z}}\leq l_{u}$,\thinspace
\thinspace\thinspace\thinspace\thinspace$l_{v_{c}}\leq l_{u}$, then

(a) for $z_{gen}>0$, it follows that
\begin{align}
0  & <z_{gen}\cdot l_{u}=-(\alpha\cdot l_{v_{z}}+\beta\cdot l_{v_{c}%
})\,\,\,\,\,\,\text{,\thinspace\thinspace\thinspace\thinspace\thinspace
\thinspace}0<z_{gen}=-(\alpha\cdot\frac{l_{v_{z}}}{l_{u}}+\beta\cdot
\frac{l_{v_{c}}}{l_{u}})\,\,\text{,}\nonumber\\
0  & <-(\alpha\cdot\frac{l_{v_{z}}}{l_{u}}+\beta\cdot\frac{l_{v_{c}}}{l_{u}%
})<-(\alpha\cdot\frac{l_{u}}{l_{u}}+\beta\cdot\frac{l_{u}}{l_{u}}%
)=-(\alpha+\beta)\,\,\,\text{,}\label{3.128}\\
\alpha+\beta & <0\,\,\,\,\,\text{,\thinspace\thinspace\thinspace
\thinspace\thinspace\thinspace\thinspace\thinspace}\alpha+\sqrt{1-\alpha^{2}%
}<0\,\,\,\,\,\,\text{,\thinspace\thinspace\thinspace\thinspace\thinspace
\thinspace\thinspace\thinspace\thinspace\thinspace\thinspace\thinspace
\thinspace}.0<\sqrt{1-\alpha^{2}}<-\alpha\,\,\,\,\text{,}\label{3.129}\\
0  & <1-\alpha^{2}<\alpha^{2}\text{\thinspace\thinspace\thinspace
\thinspace\thinspace,\thinspace\thinspace\thinspace\thinspace\thinspace
\thinspace\thinspace\thinspace\thinspace}2\cdot\alpha^{2}%
>1\,\,\,\,\text{,\thinspace\thinspace\thinspace\thinspace\thinspace
\thinspace\thinspace\thinspace\thinspace\thinspace}\alpha^{2}>\frac
12\,\,\,\,\,\text{,\thinspace\thinspace\thinspace\thinspace\thinspace}%
\mid\alpha\mid\,>\frac{\sqrt{2}}2\,\,\,\text{.}\label{3.130}%
\end{align}

(b) for $z_{gen}<0$, it follows that
\begin{align}
0  & <\alpha\cdot\frac{l_{v_{z}}}{l_{u}}+\beta\cdot\frac{l_{v_{c}}}{l_{u}%
}<\alpha\cdot\frac{l_{u}}{l_{u}}+\beta\cdot\frac{l_{u}}{l_{u}}=\alpha
+\beta\,\,\,\,\,\text{,}\label{3.131}\\
\alpha+\beta & >0\,\,\,\,\,\,\text{,\thinspace\thinspace\thinspace
\thinspace\thinspace\thinspace\thinspace\thinspace}\alpha+\sqrt{1-\alpha^{2}%
}>0\,\,\,\,\,\,\text{,\thinspace\thinspace\thinspace\thinspace\thinspace
\thinspace\thinspace\thinspace\thinspace}\sqrt{1-\alpha^{2}}>-\alpha
\,\,\,\,\text{,}\label{3.132}\\
1-\alpha^{2}  & >\alpha^{2}\,\,\,\,\text{,\thinspace\thinspace\thinspace
\thinspace\thinspace\thinspace\thinspace}1>2\cdot\alpha^{2}%
\,\,\,\,\text{,\thinspace\thinspace\thinspace\thinspace\thinspace\thinspace
}\alpha^{2}<\frac12\,\,\,\,\text{,\thinspace\thinspace\thinspace
\thinspace\thinspace\thinspace\thinspace}\mid\alpha\mid\,<\frac{\sqrt{2}%
}2\,\,\,\text{.}\label{3.133}%
\end{align}

Let\thinspace\thinspace$0<z_{gen}\leq k_{0}=-(\alpha+\beta)=-(\alpha
+\sqrt{1-\alpha^{2}})$. We can find the value of $\alpha$ for which $k_{0}$ is
a real number and $k_{0}>0$. Since
\[
k_{0}=-\alpha-\sqrt{1-\alpha^{2}}\,\,\,\,\,\,\text{,\thinspace\thinspace
\thinspace\thinspace\thinspace\thinspace\thinspace\thinspace}k_{0}%
+\alpha=-\sqrt{1-\alpha^{2}}\,\,\,\text{,}
\]
\[
k_{0}^{2}+2\cdot k_{0}\cdot\alpha+\alpha^{2}=1-\alpha^{2}\,\,\,\,\text{,}
\]
\[
\alpha^{2}+k_{0}\cdot\alpha+\frac12\cdot(k_{0}^{2}-1)=0\,\,\,\,\text{,}
\]
\begin{align*}
\alpha_{1,2}  & =\frac12\cdot(-k_{0}\pm\sqrt{k_{0}^{2}-2\cdot(k_{0}^{2}%
-1)})\,\,\,\text{,}\\
\alpha_{1,2}  & =\frac12\cdot(-k_{0}\pm\sqrt{2-k_{0}^{2}})\,\,\,\,\text{.}\,
\end{align*}

If $\alpha_{1,2}$ are real numbers then $k_{0}^{2}\leq2$ and $\mid k_{0}%
\mid\,\leq\sqrt{2}$. Therefore, if $0<z_{gen}\leq k_{0}=-(\alpha+\beta)$ then
$0<z_{gen}\leq\sqrt{2}$ and $\alpha=-(\sqrt{2}/2)$ for $k_{0}=\sqrt{2}$.

The general observed shift parameter $z_{gen}$ could not have values bigger
than $\sqrt{2}$ if $z_{gen}>0$. This means that if we could measure values of
$z_{gen}>\sqrt{2}$ the values of the general observed shift parameter could
not be explained only on the basis of the existing Doppler effect. Other
physical reasons should be taken into account if $z_{gen}>\sqrt{2}$.

\section{Conclusion}

In the present paper we have considered the notion of null (isotropic) vector
field in spaces with affine connections and metrics. On the basis of the
notion of centrifugal (centripetal) and Coriolis' velocities the notions of
standard (longitudinal) and transversal Doppler effects are introduced and
considered in spaces with affine connections and metrics. On the other side,
by the use of the Hubble law, leading to the introduction of the Hubble effect
and the aberration effect, some connections between the kinematic
characteristics of the relative velocity and the Doppler effects, the Hubble
effect, and the aberration effect are investigated. It is shown that the
Hubble effect and the aberration effect are corollaries of the standard and
transversal Doppler effects. The Hubble effect and the aberration effect could
influence each other and a general effect as a results of both effects could
be considered. The upper limit of the general observed shift parameter
$z_{gen}$ if both the effects appear is estimated at $z_{gen}=\sqrt{2}$. This
means that values of the general observed shift parameter bigger than
$\sqrt{2}$ and found experimentally could not be explained only on the basis
of the existing Doppler effects. In such cases, other physical reasons should
be taken into account.

The Doppler effects, the Hubble effect, and the aberration effect are
considered on the grounds of purely kinematic considerations. It should be
stressed that the Hubble and the Coriolis functions $H$ and $H_{c}$ are
introduced on a purely kinematic basis related to the notions of relative
velocity and to the notions of centrifugal (centripetal) and Coriolis'
velocities. Its dynamic interpretations in a theory of gravitation depends on
the structures of the theory and the relations between the field equations and
both the functions. In this paper it is shown that notions the specialists use
to apply in theories of gravitation and cosmological models could have a good
kinematic grounds independent of any concrete classical field theory. Doppler
effects, Hubble's effect, and aberration's effect could be used in mechanics
of continuous media and in other classical field theories in the same way as
the standard Doppler effect is used in classical and relativistic mechanics.

\end{document}